\documentclass[journal]{IEEEtran}
\usepackage{amsmath}
\usepackage{color}
\usepackage{amsthm} 
\usepackage{amssymb}
\usepackage{float}
\usepackage{nccmath}
\usepackage{graphicx}
\usepackage[linesnumbered,ruled]{algorithm2e}
\usepackage{setspace}
\usepackage{booktabs}
\usepackage{makecell}
\usepackage{subcaption}
\usepackage{mwe}
\usepackage{cite}
\usepackage{booktabs, multirow}
\usepackage{amsmath,amssymb,amsfonts}
\usepackage{graphicx}
\usepackage{textcomp}
\usepackage{xcolor,cancel}
\usepackage[normalem]{ulem}

\usepackage{algorithmicx}
\usepackage{xspace}
\usepackage{bbm}
\usepackage{hyperref}
\hypersetup{
	pdfpagemode=pagewidth,
	plainpages=false,
	colorlinks,
	urlcolor=blue,
	linkcolor=red,
	citecolor=green,
	bookmarksnumbered
}

\newcommand{\BfPara}[1]{{\noindent\bf#1.}\xspace}

\begin{document}
\title{Cooperative Multi-Agent Deep Reinforcement Learning for Reliable Surveillance via Autonomous Multi-UAV Control}
\author{Won Joon Yun, Soohyun Park, Joongheon Kim,~\IEEEmembership{Senior Member,~IEEE}, MyungJae Shin,\\ Soyi Jung,~\IEEEmembership{Member,~IEEE}, David A. Mohaisen,~\IEEEmembership{Senior Member,~IEEE}, and Jae-Hyun Kim,~\IEEEmembership{Member,~IEEE}
\thanks{This research was supported by National Research Foundation of Korea (2021R1A4A1030775 for Basic Research Lab and 2019M3E3A1084054). \textit{(Corresponding authors: Joongheon Kim, Soyi Jung, Jae-Hyun Kim)}}
\thanks{W. J. Yun, S. Park, and J. Kim are with the School of Electrical Engineering, Korea University, Seoul 02841, Korea e-mails: \{ywjoon95,soohyun828,joongheon\}@korea.ac.kr.}
\thanks{M. Shin is with Mofl Inc., Daejeon 50499, Korea e-mail: mjshin.cau@gmail.com.}
\thanks{S. Jung is with the School of Software, Hallym University, Chuncheon 24252, Korea e-mail: sjung@hallym.ac.kr.}
\thanks{D. Mohaisen is with the Department of Computer Science, University of Central Florida, Orlando, FL 32816, USA e-mail: mohaisen@ucf.edu.}
\thanks{J.-H. Kim is with the Department of Electrical and Computer Engineering, Ajou University, Suwon 16499, Korea e-mail: jkim@ajou.ac.kr.}
}

\maketitle
\begin{abstract}
CCTV-based surveillance using unmanned aerial vehicles (UAVs) is considered a key technology for security in smart city environments. 
This paper creates a case where the UAVs with CCTV-cameras fly over the city area for flexible and reliable surveillance services.
 UAVs should be deployed to cover a large area while minimize overlapping and shadow areas for a reliable surveillance system.
However, the operation of UAVs is subject to high uncertainty, necessitating autonomous recovery systems. This work develops a multi-agent deep reinforcement learning-based management scheme for reliable industry surveillance in smart city applications. The core idea this paper employs is autonomously replenishing the UAV's deficient network requirements with communications. Via intensive simulations, our proposed algorithm outperforms the state-of-the-art algorithms in terms of surveillance coverage, user support capability, and computational costs.
\end{abstract}
\begin{IEEEkeywords}
Multi-agent systems, Neural networks, Surveillance, Unmanned aerial vehicle (UAV)
\end{IEEEkeywords}

\section{Introduction}\label{sec:intro}
The unprecedented demands of reliable surveillance services fueled the need for flexible and reliable network supporting systems~\cite{jcn01}. 
Due to recent advances, unmanned aerial vehicles (UAVs), or drones, began to be considered as core network devices for providing flexible and reliable network services such as mobile surveillance applications using UAVs.
Using the mobility feature of UAVs, it has been shown that UAVs are able to adaptively and dynamically update the surveillance UAVs' locations~\cite{tii1,tii2,9364754,tii3,tvt19shin,6G,9467353}.
UAVs facilitate their mobility to enable line-of-sight (LOS) CCTV-based vision-enabled surveillance services, ensuring reliable monitoring services. The on-demand deployments of surveillance UAV allows them to change their positions continuously to enlarge the area covered by surveillance UAVs. Moreover, due to the relatively low-cost and the possibility of a broad range of applications, UAVs are one promising solution for robust and flexible mobile CCTV-based surveillance. However, UAVs are generally insufficiently maintained, causing them to face higher safety risks. For example, engine shutdowns due to collisions with other aircrafts or terrain can damage the wireless base station's hardware. In addition, the on-board power of UAVs is jointly consumed by their mobility and communication support functions. To prevent unexpected battery problems, such as low battery levels, the power consumption of communication, surveillance, and mobility needs to be monitored regularly.

As such, autonomous surveillance UAVs management system is essential in imbuing more robust and resilient surveillance services into UAV-based network systems. It is essential to conduct a joint optimization of the energy consumption and enhance the reliability of the network surveillance services under the behavior uncertainty of target object's movements/deployments and neighboring UAVs. Recently, a considerable amount of work has been published on the optimization of deployment of UAVs for cellular services, including optimization-based coverage control, which is essential for surveillance, for transmit power reduction and deployment of UAVs considering various uncertainties (e.g., demands changes of users)~\cite{mozaffari2016optimal}, optimal path-planning for multiple UAVs to provide wireless services to the cell edge user based on convex relaxation technique~\cite{cheng2018uav}, and the coverage maximization with minimum number of drone base stations~\cite{kalantari2016number}.

Even though the previous works show reasonable performance in terms of their objectives, the solution approaches are all centralized optimization problems. These approaches are impossible to yield an online (computational) solution for highly dynamic and distributed UAVs-enabled networks. To solve those problems in a distributed setting, machine learning based approaches are effective. To handle high dynamics of UAVs-enabled network (e.g, surging demands of users, malfunction of neighboring UAVs, etc.) with high uncertainty and dynamic updates, a new multi-agent deep reinforcement learning (DRL) scheme is designed in this study, for distributed computation over UAVs, considering users and multiple UAVs.

Compared to the conventional optimization approaches, various ML techniques have been applied to improve the performance of UAV-based computing, including an ML-based approach for autonomous trajectory optimization~\cite{chen2017learning} and the optimization of UAV location in a downlink system with a joint K-means and expectation maximization (EM) based on a Gaussian mixture model (GMM)~\cite{lu2020machine}, dynamic optimization of the locations of UAVs in a VLC-enabled UAV based network for minimizing the transmit power~\cite{wang2019deep}, among others.
There are some studies that utilize DRL methods in UAV network systems, including the meta-reinforcement learning-based path-planning for UAVs in dynamic and unknown wireless network environments~\cite{hu2020meta}, a Q-learning method-based dynamic location planning of UAVs in a non-orthogonal multiple access (NOMA) based wireless network~\cite{liu2019uav}, and the optimization for UAV optimal energy consumption control considering communication coverage, fairness, and connectivity.

However, these general ML approaches cannot be applied to deterministic multi UAVs decision-making under uncertain environments, leading to undesirable outputs. This is mainly because the aforementioned studies do not consider the UAV-based network system's partially observable multi-agent environment, since each agent has different information, and  UAVs that cannot fully exchange information. Moreover, these approaches only focus on the network communication system even if visual information is available.
This underlines the need for more advanced research in the communication between multiple surveillance UAVs.
Thus, this paper builds on the existing issues by creating a novel algorithm that exchanges partially observed information from a particular UAV to other UAVs in an uncertain and constantly varying environment.

\begin{figure*}[t!]
    \centering 
    \includegraphics[width=0.9\linewidth]{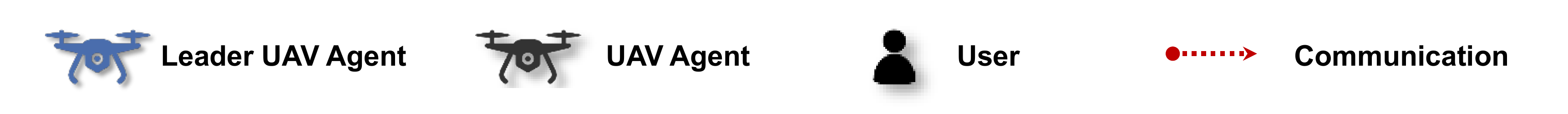}\\
    \includegraphics[width=0.9\linewidth]{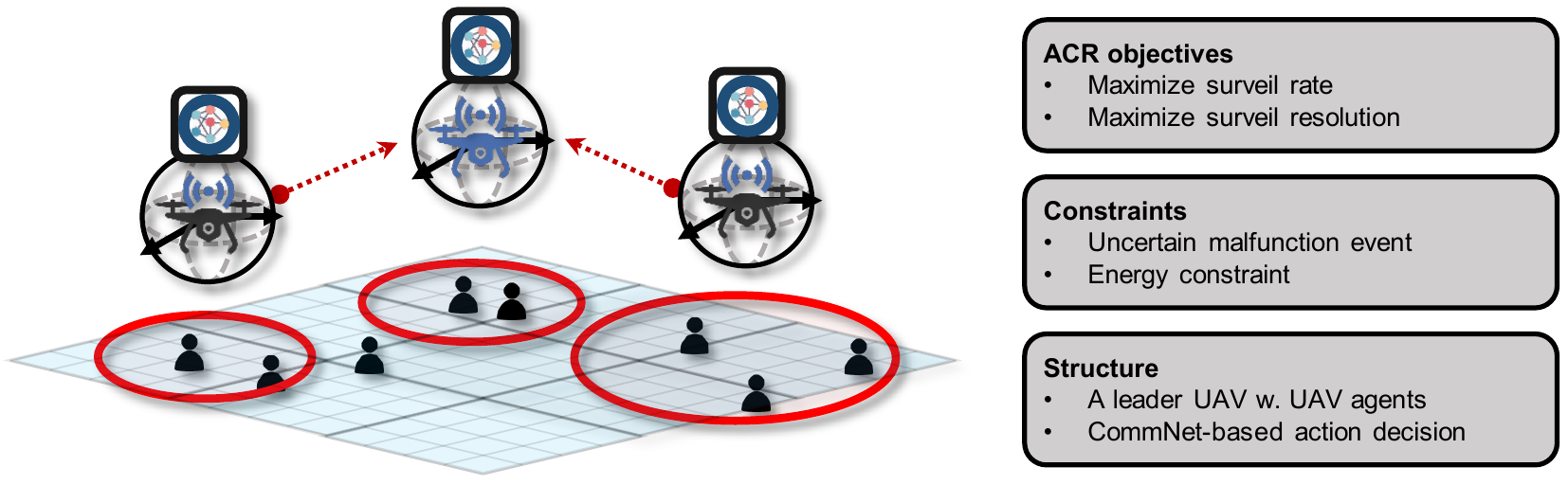}
    \caption{\textit{CommNet}-based autonomous and reliable surveillance management systems for smart city applications.}
    \label{fig:overview}
\end{figure*}

Fig.~\ref{fig:overview} shows a schematic system that consists of two units: the UAVs, and the target of surveillance, which is called the \textit{user}. Some UAVs cooperatively exchange information for reliable monitoring services, while uncooperative UAVs do not.
The objective of the surveillance UAV is to autonomously maneuver itself to an area with the highest number of users. Hence, it is essential to configure a system where UAVs automatically induce the optimal trajectories and coverage to achieve this objective. In this process, the autonomous optimization system needs to take into account the characteristics of UAVs (e.g., on-board battery). This paper configures the autonomous surveillance resilience system based on a multi-agent DRL (MADRL) scheme, called communication neural network (\textit{CommNet})~\cite{sukhbaatar2016learning}. The goal of the proposed scheme is to optimize the energy consumption of UAVs and trajectories that strengthen surveillance performance, which means deploying the surveillance UAVs to cover the area with the most number of users.
The monitoring of the target is achieved through a camera lens attached to the body of the surveillance UAV. Hence, this paper also analyzes the relationship between the surveillance resolution and its corresponding area covered. The surveillance resolution is defined as the resolution of the images captured by the surveillance UAVs. It is observed through the experiments that the lower the resolution of the captured image, the larger the area covered by that particular UAV.

To the best of the author's knowledge, the autonomous UAVs cooperative surveillance management system (under the consideration of UAVs that do not fully exchange information) via MADRL has not been studied yet. Thus the proposed scheme will guide dynamic UAV-based autonomous surveillance studies in the future. To sum up, the contributions of this paper are as follows:
\begin{itemize}
    \item This paper proposes a new MADRL algorithm to achieve autonomous UAV cooperation in a distributed UAV context {to estimate} the uncertainty of the environment and optimize energy consumption and overall surveillance reliability and operation.
    \item Experiments conducted in this paper show the relationship between the performance of the surveillance UAVs, which refers to the resolutions of the captured surveillance UAV data, and the area covered by the corresponding surveillance UAV.
    \item Through performance evaluations of the proposed \textit{CommNet} based autonomous UAV surveillance management scheme, this work shows the scheme's ability to successfully perform reliable and robust cooperative management of distributed UAVs. 
\end{itemize}

The rest of this paper is organized as follows. Sec.~\ref{sec:rl} summarizes the state-of-the-art of reinforcement learning algorithms. Sec.~\ref{sec:method} presents the proposed autonomous UAV coordination inspired by \textit{CommNet}-based reinforcement learning algorithms. Sec.~\ref{sec:ex} evaluates the performance of the proposed autonomous UAV coordination algorithm. Finally, Sec.~\ref{sec:con} concludes this paper and provides future research directions.

\section{Deep Reinforcement Learning}\label{sec:rl}
\subsection{Preliminaries}\label{subsec:pre}
In this work, a Markov decision process (MDP) is defined as a tuple $(\mathcal{S}, \mathcal{A}, P, R, T)$, where $\mathcal{S}$ is a finite set of all valid states, and $\mathcal{A}$ is a finite set of all valid actions. The function $P : \mathcal{S} \times \mathcal{A} \to P(\mathcal{S})$ is a transition probability function, with $P(s'\mid s, a)$ being the probability of transitioning into state $s'$ if an agent starts executing action $a$ in state $s$. The function $R : \mathcal{S} \times \mathcal{A} \times \mathcal{S} \to \mathbb{R}$ denotes the reward function, with $r_{t} = R(s_t, a_t, s_{t+1})$. The MDP has a finite time horizon $T$, and solving an MDP means finding a policy $\pi_\theta  \in \Pi : \mathcal{S} \times \mathcal{A} \to \left[ 0, 1 \right]$, where $\pi$ is parameterized with $\theta$ (e.g., the weights and biases of a neural layer); $\pi_\theta$ specifies which action $a \in \mathcal{A}$ must be executed in each $s$ for maximizing the discounted cumulative rewards received during the finite time $T$. 

When the environment transitions and the policy are stochastic, the probability of a $T$-step trajectory is defined as $P(\tau \mid \pi_{\theta}) = \rho(s_0)\prod_{t=0}^{T-1}P(s_{t+1}\mid s_t,a_t)\pi_{\theta}(a_t \mid s_t)$
where $\rho$ is the starting state distribution. Then, the expected return $\mathcal{J}(\pi_{\theta})$ is defined as $\mathcal{J}(\pi_{\theta}) = \int_{\tau}P(\tau \mid \pi_{\theta})R(\tau) = \mathbb{E}_{\tau \sim \pi_{\theta}}\left[R(\tau)\right]$
where the trajectory $\tau$ is a sequence of states and actions in the environment. 
The goal in reinforcement learning is to learn a policy that maximizes the expected return $\mathcal{J}(\pi_{\theta})$ when the agent acts according to the policy $\pi_{\theta}$. Therefore, the optimization objective is expressed by
\begin{equation}
    \label{eq:obj_rl}
\pi_{\theta}^* = \arg\max_{\theta} \mathcal{J}(\pi_{\theta})
\end{equation}
with $\pi_{\theta}^*$ being the optimal policy. The multi-agent MDP (MMDP) \cite{boutilier1996planning} generalizes the MDP to the multi-agent system, where the state space is defined by taking the Cartesian product of the state spaces of the individual agents, and actions represent the joint actions that can be executed by the agents. 

\subsection{Deep reinforcement learning}\label{subsec:rl}
\BfPara{Deep \textit{Q}-Network (DQN)\cite{mnih2013playing}}
A deep \textit{Q}-network (DQN) is a model-free reinforcement learning method designed to learn the optimal policy with a high-dimensional state space. The DQN is inspired by \textit{Q}-learning\cite{watkins1992q}, where a neural network is used to approximate the \textit{Q}-function. Experience replay $\mathcal{D}$ and target network are two key features used to stabilize the optimization. Experiences $e_t = (s_{t}, a_{t}, r_{t+1}, s_{t+1})$ of the agent are stored in the experience buffer $\mathcal{D} = (e_1, e_2, \dots, e_T)$, and are periodically resampled to train the $\textit{Q}$-networks. A mini-batch resampled experience is used to update the parameters $\theta_i$ of the policy with the loss function at the $i$-th  training iteration where the loss function is defined as
\begin{multline}
    \label{eq:dqn}
    L(\theta_i) = \mathbb{E}\left[(r_{t+1} + \right.\\ \left.\gamma \max_{a'}Q(s_{t+1}, a' ; \theta_{i}^{-}) - Q(s_t, a_t ; \theta_{i}))^2\right]
\end{multline}
where $\theta^{-}_i$ are the target network parameters. The target network parameters $\theta^{-}_i$ are updated using the \textit{Q}-network parameters $\theta$ in every predefined step. The stochastic gradient descent method is used to optimize the loss function. 

\BfPara{Proximal Policy Optimization (PPO)\cite{schulman2017proximal}}
PPO is one of the breakthroughs of DRL algorithms, which adopts the trust region concept~\cite{schulman2015trust} to improve the training stability by ensuring that $\pi_\theta$ updates at every iteration are small by clipping the probability ratio $r_{\pi}(\theta) = \pi_\theta(a \mid s) / \pi_{\theta_{old}}(a \mid s)$, where $\theta$ is the parameter of the new policy while $\theta_{old}$ is that of the old policy. A clipped surrogate objective to prevent the new policy from straying away from the old one is used to train the policy $\pi_\theta$. The clipped objective function is as follows:
\begin{equation}
    \label{eq:ppo_surrogate}
    L^{\text{CLIP}}_t(\theta) = \min(r_t(\theta)A_t, \textit{clip}(r_t(\theta),1-\epsilon,1+\epsilon)A_t),
\end{equation}where $A_t$ is the estimated advantage value under hyperparameter $\epsilon < 1$, which denotes how far away the new policy is allowed to update from the old policy. PPO uses the stochastic gradient descent to maximize the objective \eqref{eq:ppo_surrogate}. 

\BfPara{Limitations and MADRL}
The prior approaches are designed with a single agent system. In this system, the agent considers only changes that are the outcomes of its own actions. However, in a multi-agent system, the agent needs to concurrently observe the effect of its own actions as well as the behavior of other agents. This characteristic of the multi-agent system constantly reshapes the environment and leads to non-stationarity (i.e., lead to a non-stationary problem). As a result, the convergence theory of the predecessors is generally not guaranteed in multi-agent systems~\cite{nguyen2018deep}. Therefore, the information collection and processing method should not affect the convergence stability of the agents in multi-agent systems. 
Thus, various MADRL algorithms are proposed for multi-agent cooperation and coordination. 
In centralized MADRL algorithms, multi-agent states and actions are formulated as the inputs and outputs of a single deep neural network. Thus, there is only one policy that determines MADRL actions. In addition, the centralized policy collects all the observable information (fully observable information) and determine{s} the actions of all agents at once. 
In \textit{CommNet}~\cite{CommNet,icte21yun,icoin21jung} which is one of well-known centralized MADRL algorithms, the cooperation and coordination among multiple agents is formulated by mixing information in each hidden layer computation.   

\section{Autonomous Multi-UAV Coordination for Reliable Surveillance}\label{sec:method}

\subsection{System description}
Suppose that $N$ users, $M$ multi-agent cooperative agents, {$K$ non-agent UAVs} are deployed. 
The sets of users, UAVs agents, and {non-agent UAVs} are denoted as $\mathcal{U} \triangleq \left\{ u_{1}, u_{2}, \dots, u_n, \dots, u_{N} \right\}$, $\mathcal{B} \triangleq \left\{ b_{1}, b_{2}, \dots, b_m, \dots, b_{M} \right\}$ {and $\mathcal{H} \triangleq \left\{ h_{1}, h_{2}, \dots, h_k, \dots, h_{K} \right\}$}, respectively.
In addition, $u_n$, $b_m$, and {$h_k$} represent $n$-th user, $m$-th UAV agent,  and {$k$-th non-agent UAV}, where $\forall u_n \in \mathcal{U}, n \in [1, N]$, $\forall b_m \in \mathcal{B}, m \in [1, M]$, and {$\forall h_k \in \mathcal{H}, k \in [1,K]$}. 
The proposed system assumes a leader UAV agent for handling communications between UAV agents. The leader UAV agent receives information for multi-agent cooperation, as shown in Fig.~\ref{fig:commnet}. 
In addition, this paper assumes that each user $u_n$ is associated with only one UAV $b_{m}$, whereas each UAV can be associated with multiple users. A camera is embedded on each UAV. Each UAV captures surveillance coverage $R^C$ with the resolution $q_i \in \mathcal{Q}, \forall i \in [1, I]$, where $\mathcal{Q}$ stands for the set of resolutions. This study presupposes that each camera sensor is not affected by zoom-in, zoom-out, or the UAV's movement. These assumptions are reasonable since, according to \cite{Samsung}, the camera sensor dynamically controls the sizes of micro pixels, so that the resolution is not degraded even when zooming-in.
Therefore{,} the surveillance resolutions have an inverse relationship with surveillance areas.
Each UAV $b_{m}$ can conduct multi-agent cooperation to manage its surveillance coverage autonomously. This paper assumes that the wireless connection between inter-UAVs is good enough to deliver UAV's information without losing robust and stable MADRL training. Each UAV has a maximum capacity of the battery.

The objective of the proposed autonomous coordination for reliability (ACR) is to increase the regions of monitoring areas for surveillance reliability with high resolution. Therefore, the proposed algorithm in this paper tries to achieve reliable surveillance under the scenario that the number of users to be observed changes while considering the UAV's conditions \textit{e.g.,} the possibilities that the UAVs are dropped, malfunctioned, or energy-exhausted. 
At the same time, each UAV tries to optimize energy consumption to combat the power-hungry nature of UAVs.

\subsection{MADRL formulation}
For the MADRL formulation of the proposed ACR, the state space, action space, and rewards should be defined. 

\subsubsection{State space} 
{The state space of our proposed ACR consists of location information (\textit{e.g.}, absolute position information, and the relative position or distance information with other users/UAVs), energy information, and surveillance information (\textit{e.g.}, whether user is monitored or not; and which UAV monitors with which resolutions).
The absolute positions of each UAV agent, user, and non-agent UAV are defined as $p(b_m)$,$\forall m \in [1, M]$, $p(u_n)$, $\forall n \in [1, N]$, and $p(h_k)$, $\forall k \in [1, K]$, respectively. 
The relative positions for $b_m$ with $b_{m'}$, $u_n$, and $h_k$ are denoted by $l(b_m, b_{m'})$, $l(b_m, u_n)$, and $l(b_m, h_k)$, respectively. The distance for $b_m$ with $b_{m'}$, $u_n$, and $h_k$ are denoted as $d(b_m, b_{m'})$, $d(b_m, u_n)$, and $d(b_m, h_k)$. Note that $l(b_m,(\cdot))$, and $d(b_m,(\cdot))$ stand for the relative position and physical distance between $b_m$ and $(\cdot)$, respectively. The location information of $b_m$ is defined as $\mathbf{p}_m \triangleq \left\{ p(b_m), \bigcup_{x} \{l(b_m, x), d(b_m,x)\}\right\}$, where $x \in \{\mathcal{B}\cup\mathcal{U}\cup\mathcal{H}\}\backslash b_m$.}
In each time, every single UAV consumes basic operational energy for aviation and monitoring, where the basic energy consumption of $b_m$ is denoted by $e^b_m$. In addition, $e^c_m$ denotes the energy consumption of $b_m$, which depends on the surveillance coverage range; and it can be computed as follows:
\begin{eqnarray}
    e^b_m &=& (\delta + \zeta\times h)\times t + P(h/v), \label{eq:ep}\\
    e^c_m &=& R^C_m \times  \rho_{R^C}.\label{eq:ec}
\end{eqnarray}
In \eqref{eq:ep}, $\delta$ is the minimum power consumption to fly UAV over the ground, and $\zeta$ is the speed multiplier of the motor operation. 
Here, $v$ and $t$ represent the speed and  operating time. The required power consumption for lifting the UAV to height $h$ at speed $v$ is denoted as $P(h/v)$~\cite{pugliese2016modelling, zorbas2016optimal}. 
In \eqref{eq:ec}, $R^{C}_m$ and $\rho_{R^C}$ represent the current coverage range of the $m$-th UAV agent and the coefficient of monitoring energy consumption. Note that the monitoring related energy consumption is related with the radiation and signal processing. {Here, the energy information of $b_m$ is defined as $\mathbf{e}_m \triangleq \{e^b_m, e^c_m\}$.}
{$R^{C}_m$ depends on the surveillance resolution of $b_m$, which is denoted by $q(b_m) \in \bigcup^{I}_{i=1}\{q_i\}$. The surveillance coverage of $b_m$ is defined as positional set and is written as follows:
\begin{equation}
\Lambda_m \triangleq \{p(\psi) | d(b_i,\psi) \leq R^C_m\}, \forall \psi \in \Psi,
\end{equation}
Note that $\Psi$, and $\psi$ represent surveilance field and arbitrary point on $\Psi$, respectively.  
Whether $b_m$ surveils $u_n$ is determined with the allocation variable $c_{mn}$, which follows the surveillance rule defined as,}
\begin{align}
    \sum^{M}_{m=1}c_{mn}\leq 1 & , \tag{C1} \\
    c_{mn} \leq \mathbbm{1}(q_m &\leq \max_{m'}(q(b_{m'})))\cdot \mathbbm{1}(b_m \in \cup_{m'}\{b_{m'}\}), \tag{C2} \\
    \forall c_{mn} \in \{0,1&\}, \forall b_{m'} \in \mathcal{B}\backslash \{b_m|d(b_m,u_n) > R^C_m\}. \tag{C3}
\end{align}
{Note that $\mathbbm{1}(\cdot)$ is indicator function. Here, the surveillance information of $b_m$ is defined as $\mathbf{c}_m\triangleq\{R^C_m, q(b_m), \Lambda_m,\bigcup_{n}^{N}\{c_{mn}\}\}$.}
Taking all the above into consideration, the set of states is denoted as {$\mathcal{S} \triangleq \left\{s_{1}, s_{2}, \dots, s_{m}, \dots, s_{M} \right\}$ where $s_{m}$ represents the state of $m$-th UAV which is defined as $s_{m}\triangleq \left\{\mathbf{p}_m, \mathbf{e}_m, \mathbf{c}_m \right\}$.}  

\begin{figure}[t!]
    \centering
    \includegraphics[width=0.45\columnwidth]{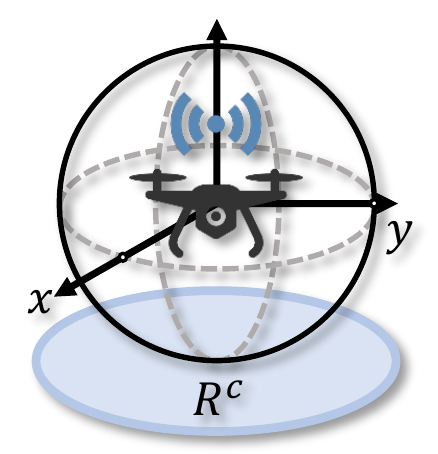}
    \caption{UAV coordination}
    \label{fig:coordination}
\end{figure}

\subsubsection{Action space} 
The action space of the proposed ACR consists of $10$ discrete actions, i.e., $8$ actions are for moving actions $\pm \alpha$ in $x$ direction, $\pm \alpha$ in $y$ direction, or $\pm \beta$ in both $x$ and $y$ direction; and the other $2$ actions are for controlling the surveillance resolution level $\pm 1$, where current position is $(x,y)$ and surveillance coverage range is $R^{C}$.
The illustrative description for this action space is as shown in Fig.~\ref{fig:coordination}.
To achieve monitoring service, each UAV takes moving actions and coverage range controlling actions. The discrete set of actions is defined as $\mathcal{A} = \{x_{pos} + \alpha, x_{pos} - \alpha, y_{pos} + \alpha, y_{pos} - \alpha,
(x_{pos} + \beta, y_{pos} + \beta),(x_{pos} + \beta,y_{pos} - \beta), (x_{pos} - \beta, y_{pos} + \beta), (x_{pos} - \beta, y_{pos} - \beta), q_{i-1}, q_{i+1} \}$.

\subsubsection{Reward}

The rewards of the proposed ACR are classified into two groups, i.e., UAV rewards and cooperation rewards.

\BfPara{UAV Reward}
For defining the rewards in UAVs, this paper considers energy consumption, battery discharge, and the number of users. 
The UAV is power-hungry by nature, the rewards for $m$-th UAV with energy consumption $r^c_m$, and the $r^c_m$ of each UAV agent is the scaled summation of energy consumption of $e^{b}_{m}$ and $e^{c}_{m}$:
\begin{equation}
    \label{eq:reward_consumption}
    r^e_m = -\rho_{e1} \times \left(e^{b}_{m} + e^{c}_{m}\right) - \rho_{e2} \times \mathbbm{1}(b_m \mathrm{~is~operating}), 
\end{equation}
where $\rho_{e1}$, and $\rho_{e2}$ are the scaling factor for energy consumption and for aviation status of UAV agent, respectively.

The reward for the surveillance $r^c_m$ is defined as the ratio of the number of users in the current surveillance coverage to the number of users when the surveillance coverage is maximized. The surveillance reward for $b_m$ is written as follows:

\begin{equation}
    r^c_m = \rho_c \times \mathrm{log}(q(b_m)) \times \sum^N_{n=1} c_{mn},
\end{equation}
where $\rho_c$ is the scaling factor for surveillance.
As $r^c_m$ is increased, the surveillance resolution and the number of monitored/observed users are also increased. That is obviously beneficial in terms of surveillance reliability. 

\BfPara{Cooperation Reward}
To define the rewards for the cooperation among UAVs, this paper addresses the overlapped area among UAVs, and the number of users. 
If there are a lot of overlapped areas {regarding} surveillance coverage, it is obviously not good in terms of energy and resource efficiency. Thus, {the} reward formulation also aims to minimize overlapped areas. 
The overlapped UAV agent surveillance coverage is computed as:
\begin{equation}
    \label{eq:overlapped}
\omega = 1- \frac{\sum^M_{m=1}\left[S(\Lambda_m \cap_{m'\neq n}\{\Lambda_{m'}\}^C)\right]}{S(\bigcup^M_{m=1}\{\Lambda_m\})},
\end{equation}
where $S(\cdot)$ returns the area of input. 
This paper proposes the \textit{overlapped threshold}, $\omega^{th}$, for reward constraint of the overlapped area among UAVs. 
The reward of the total utilization $r^u$ is defined as the number of users who use the service and total number of users, and it is computed as follows:

\begin{equation}
    \label{eq:reward_overlapped}
r^u= 
\begin{cases}
 \rho_u\times\frac{1}{N}\sum^{M}_{m=1}\sum^{N}_{n=1}c_{mn}, & \mbox{if } \omega < \omega^{th} \\
 0,  & \mbox{otherwise.} 
\end{cases} 
\end{equation}
where $\rho_u$ is a scaling factor for total utilization.
Therefore, the total reward for each agent $r_{m}^{total}$ is defined as follows:
\begin{equation}
    \label{eq:tot_reward}
    r^{total}_{m} = \underbrace{r^{e}_{m}+r^u_m}_{\textrm{UAV Reward}} + \underbrace{r^c}_{\textrm{Cooperation Reward}}
\end{equation} 
\subsection{Algorithm for learning cooperation}

\begin{algorithm}[t]
\small
    Initialize the \textit{critic} and \textit{actor} networks with weights $\theta^{Q}$ and $\theta^{\mu}$ \\
    Initialize the target networks as: $\theta^{\hat{Q}} \leftarrow \theta^{Q}, \theta^{\hat{\mu}} \leftarrow \theta^{\mu}$ \\
    \For{episode = 1, MaxEpisode}{
            $\triangleright$ Initialize \textbf{UAV Agent Environments}\\
            \For{time step = 1, T}{
            $\triangleright$ With probability $\mu(\mathcal{S}|\theta^{\mu})$ select a set of actions $\mathcal{A}$ for each $s^{m}_{t} \in \mathcal{S}$\\ 
            $\triangleright$ Execute actions at in \textbf{Simulation Environments} and observe reward $\mathcal{R}_{total}$ and the next set of states $\mathcal{S}'$\\
            $\triangleright$ Store the transition pairs $\xi = (\mathcal{S}, \mathcal{A}, \mathcal{R}_{total}, \mathcal{S}')$ in replay buffer $\Phi$\\
            \textbf{If} \textit{time step} \textbf{is update period, do followings:}\\
            $\triangleright$ Sample a random minibatch from $\Phi$\\
            $\triangleright$ Set $y_{i} = r_{i} + \delta \hat{Q}(s_{i}^{'}, \mu(s_{i}^{'}|\theta^{\hat{\mu}})|\theta^{\hat{Q}})$ \\
            $\triangleright$ Update the $\theta^{Q}$ by applying stochastic gradient descent to the loss function of \textit{critic} network: $L = \frac{1}{\varphi}\sum_{i}{(y_{i} - Q(s_{i}, a_{i}|\theta^{Q}))}^{2}$ \\
            $\triangleright$ Update the $\theta^{\mu}$ by applying stochastic gradient ascent with respect to the gradient of \textit{actor} network: \\
            $\nabla_{\theta^{\mathcal{A}}}J(\theta^{\mu}) \approx \frac{1}{\varphi}\sum_{i}{\nabla_{a}Q(s, a|\theta^{Q})\nabla_{\theta^{\mu}}\mu(s|\theta^{\mu})|_{s=s_{i},a=\mu(s_{i}|\theta^{\mu})}}$ \\
            $\triangleright$ \textit{Target Update} $\theta^{\hat{Q}}$ and $\theta^{\hat{\mu}}$
            }
        }
    \caption{Autonomous multi-UAV coordination for reliable surveillance}
    \label{alg:commnet}
\end{algorithm}

\begin{figure}[t!]
    \centering
    \includegraphics[width=0.69\columnwidth]{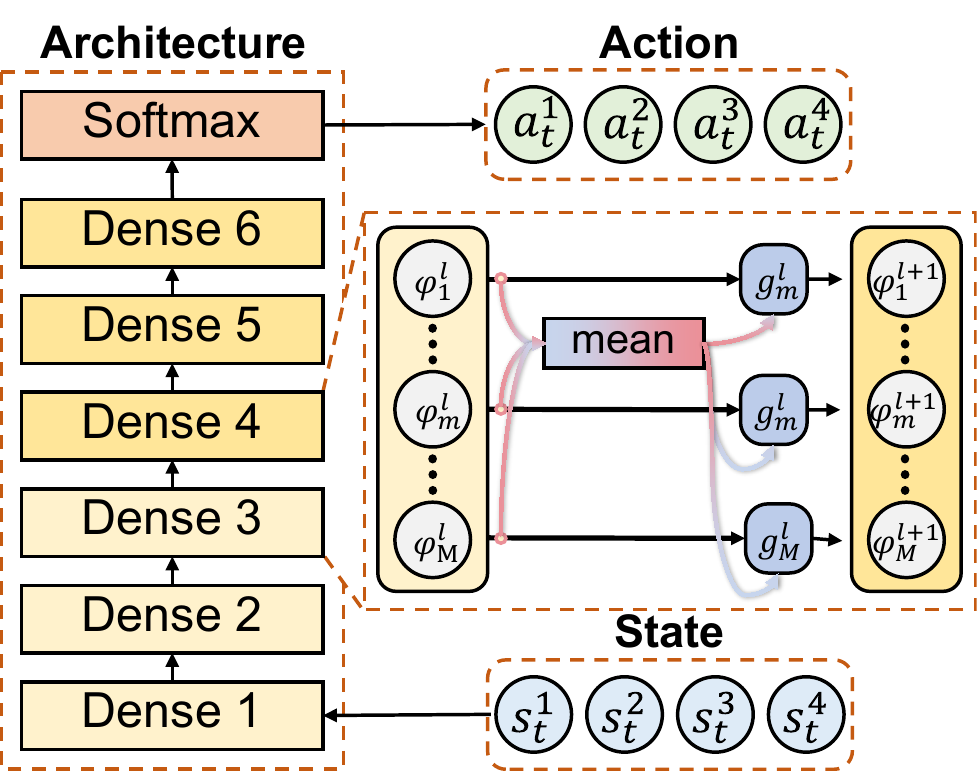}
    \caption{Structure of \textit{CommNet}.}
    \label{fig:commnet}
\end{figure}

The UAV agents of ACR exchange pieces of information of users and other UAVs. With communications UAVs and other UAV agents, the autonomous UAV coordination scheme tries to learn how to maintain the surveillance under environment uncertainty, which is shown as Algorithm \ref{alg:commnet}. 
The proposed ACR scheme considers multi-agents system, so \textit{CommNet}, a representative communication based multi-agent DRL algorithm, is applied. In \textit{CommNet}, each agent uses only its observable state as shown in Fig. \ref{fig:commnet}. 
The \textit{CommNet} enables communications among multiple agents using a single deep neural network. 
Note that conventional \textit{CommNet} considers a central server that collects pieces of information of agents and distributes processed information. 
That is, each agent has access to a central server to share information. However, the proposed scheme configures one of the UAV agents that is randomly selected as a leader UAV, and the leader collects and distribute information, i.e., other agents' information. Except the leader UAV, each UAV agent sends its embedded state information as the communication message to the leader UAV. 
The leader UAV collects the embedded information, and averages all of the received embedded messages. 
After that, the averaged message is taken as the input of the next layer. For other UAV agents, the leader UAV sends the averaged message to other UAV agents. The final layer output determines the agent's action. Here, $l \in \left\{1,..,L\right\}$, where $L$ is the number of communication steps. Each $g^l$ takes two input vectors for each UAV agent $b_m$: the hidden state $\varphi^{l}_{m}$ and the communication $comm^{l}_{m}$. The output of $g^l$ is a vector $\varphi^{l+1}_{m}$. In \textit{CommNet}, the communication and hidden state are calculated as follows:
\begin{eqnarray}
    \label{eq:hidden}
    \varphi^{l+1}_{m} &=& g^{l}(h^{l}_{m}, comm^{l}_{m}), \\ 
    \label{eq:comm}
    comm^{l}_{m} &=& \frac{1}{M-1} \sum_{m'\neq m}\nolimits \varphi^{z}_{m'}.
\end{eqnarray}
  The softmax activation function is placed at the output layer $output = \textit{softmax}(\varphi^{L}_{m})$. Then, the output of the softmax can be interpreted as the probabilities that action $a_m$ is taken when the UAV agent $b_{m}$ observes state $s_m$. This paper adopts the \textit{actor} and \textit{critic} reinforcement learning model based \textit{CommNet}~\cite{a2c}.
The overall training process is then defined as follows:
\begin{enumerate}
    \item For \textit{actor} and \textit{critic} networks, the weight parameters, i.e., $\mu$ and $Q$, are initialized (line 1).
    \item The weight parameters of the target \textit{actor} and \textit{critic} networks, i.e., $\hat{\mu}$ and $\hat{Q}$, are initialized (line 2).
    \item The set of UAVs $\mathcal{B}$ recursively follows this procedures for learning autonomous UAV coordination policies: 
    (i) For every episode, the transition pairs $\xi =\left(\mathcal{S}, \mathcal{A}, \mathcal{R}_{total}, \mathcal{S}'\right)$ are stored in \textit{replay buffer} $\Phi$. Here, $\mathcal{S}$, $\mathcal{A}$, $\mathcal{R}_{total}$, and $\mathcal{S}'$ stand for the set of states, the set of actions generated by the parameter of \textit{actor} $\theta^{\mu}$, the reward of UAV agents, and the observed set of next state spaces. Note that all transition pairs are derived from ACR environments (lines 4--8).    
    (ii) When it comes to update period, a minibatch is randomly sampled from $\Phi$. Using \textit{Bellman optimality equation} (line 11). The $i$-th transition pair of the minibatch, mean squared Bellman error is calculated with target value $y_{i}$ and $Q(s_{i}, a_{i}|\theta^{Q})$ to update the \textit{critic} network (line 12)~\cite{sutton1997significance}. The parameters of the \textit{actor} network, i.e., $\theta^{\mu}$, are updated via gradient-based optimization (line 13--14).     
    (iii) After updating the parameters of \textit{actor} and \textit{critic}, i.e.,  $\theta^{\mu}$ and $\theta^{\mathcal{Q}}$, the target parameters $\theta^{\hat{\mathcal{Q}}}$ and $\theta^{\hat{\mu}}$ are updated (line 15).
    \item The parameters of \textit{actor} and \textit{critic} are shared for UAV agents; thus the UAV agents have the same parameters of cooperation policy. Note that this sharing procedure ensures the easy usages of more UAV agents.
\end{enumerate}

\section{Performance Evaluation}\label{sec:ex}
\subsection{Performance metric}
\begin{table}[t!] 
\scriptsize
\small
\caption{Environmental setup parameters.}
\centering
\begin{tabular}{@{}ll@{}}\toprule
\multicolumn{1}{c}{\centering\textbf{Parameters}} & \multicolumn{1}{c}{\centering\textbf{Value}}\\  \midrule
\multicolumn{1}{l}{Energy for hovering, ${e^b}_{\{v=0\mathrm{km/h}\}}$} & \multicolumn{1}{c}{128.89 W} \\
\multicolumn{1}{l}{Energy for flying, ${e^b}_{\{v=20\mathrm{km/h}\}}$} & \multicolumn{1}{c}{170.32 W} \\
\multicolumn{1}{l}{Energy for surveillance, ${e^c}$} & \multicolumn{1}{c}{5 W} \\
\multicolumn{1}{l}{Time steps per episode, $T$} & \multicolumn{1}{c}{{40} min}\\ 
\multicolumn{1}{l}{Surveillance resolution, $\mathcal{Q}$} & \multicolumn{1}{c}{$[720, 1080, 2160]\mathrm{p}$}\\ 
\multicolumn{1}{l}{The number of UAV agents, $M$} & \multicolumn{1}{c}{4}\\
\multicolumn{1}{l}{The number of users, $N$} & \multicolumn{1}{c}{25}\\
\multicolumn{1}{l}{{The number of non-agent UAVs, $K$}} & \multicolumn{1}{c}{{3}}\\ 
\multicolumn{1}{l}{Overlapped threshold, $\omega_{th}$} & \multicolumn{1}{c}{0.5} \\
\toprule
\end{tabular}
\label{parameter}
\vspace{-10pt}
\end{table}
In {ACR} dynamic environment, the number of monitored users continuously change. 
{This study investigates} the convergence of {ACR} and its benchmark schemes. 
The UAVs transfer the latest locations. Every episode starts with randomly spreading agent UAVs on the grid. Each UAV is then randomly assigned a specific area of interest which is the last location left by UAV. In order to provide an optimal surveillance service, which is the common goal of agent UAVs, they must provide the best surveillance for each area. 
As a benchmark, this paper compares our proposed CommNet-based ACR to state-of-the-art techniques as follows:
\begin{enumerate}
    \item ACR with communication (Comp1): In Comp1 scheme, all agents collect observation information of others. Comp1 ACR utilizes the state-of-the-art algorithm (\textit{communication learning}). In the comparison experiment, this paper compares the performance and computational cost. In addition, this work analyzes the efficiency of the proposed scheme in Sec.~\ref{evaluation}. Note that all agents in Comp1 make action decision with CommNet-based policy.
    \item ACR with disconnection (Comp2): In Comp2 scheme, there is no leader UAV agent which collects observation information of other agents. Comp2 ACR utilizes the state-of-the-art algorithm (\textit{independent actor-critic} (IAC)-based algorithm). This work compares the proposed scheme to Comp2 corresponding to cooperation between inter-agent. Note that the structure of policy in Comp2 ACR (i.e., DNN-based policy) is equivalent to CommNet-based policy without~\eqref{eq:hidden} and~\eqref{eq:comm}. 
\end{enumerate}

\subsection{Simulation setup}
 In order to numerically analyze the performance of the proposed ACR scheme, the paper considers a $2$-dimensional {$2,400\times2,400~[\mathrm{m}^2]$} grid map, $N=25$ randomly distributed users, and $M=4$ multi-agents system based UAV agents. This experiment setup assumes each episode has a total of $40$ time steps. The initial positions of agent UAVs are set to center of the grid. In addition, the initial positions of non-agent UAVs are detached  750m from the center of the grid. Each non-agent UAV is randomly malfunctioned with probability 0.03 for each step. Note that ACR parameters are summarized in Table~\ref{parameter}. In addition, experiments were conducted with the system parameters $\alpha=333$m, $\beta = 236$m, $\sigma = 2$, $\rho_{e1}=1$, $\rho_{e2}=1$, $\rho_c=1$, and $\rho_u=3$.
This study configured two neural network structures (\textit{e.g.}, CommNet and DNN) of ACR as follows. The neural networks consist of six dense layers. In the first layer to the last layer, the number of nodes is {$64$}. The ReLU function is used in first layer to fifth layer, respectively. A Xavier initializer is used for weight initialization. This paper uses  Adam optimizer with the learning rate $0.001$. In the training procedure, $\epsilon$-greedy method is used to make the UAV agents experience a variety of actions. The initial value of epsilon is $0.3$ and the annealing epsilon is $0.0001$. 

\subsection{Evaluation results}\label{evaluation}
In the following, the study investigates the reward convergence, surveillance, trained behaviors of the proposed scheme, and comparison schemes, and computation cost comparison.
\begin{figure}[t!]
\centering
\setlength{\tabcolsep}{2pt}
\renewcommand{\arraystretch}{0.2}
\begin{tabular}{c}
\includegraphics[page=1, width=0.6\columnwidth]{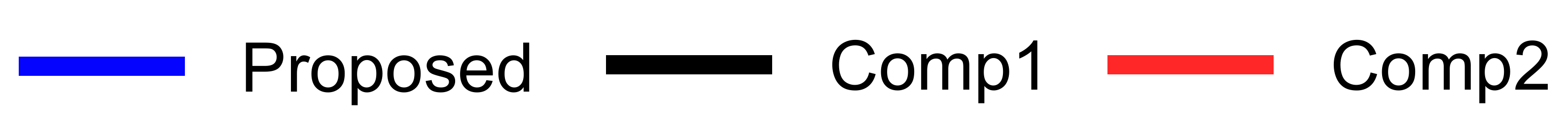}
\end{tabular}
\begin{tabular}{cc}
\includegraphics[page=1, width=0.45\columnwidth]{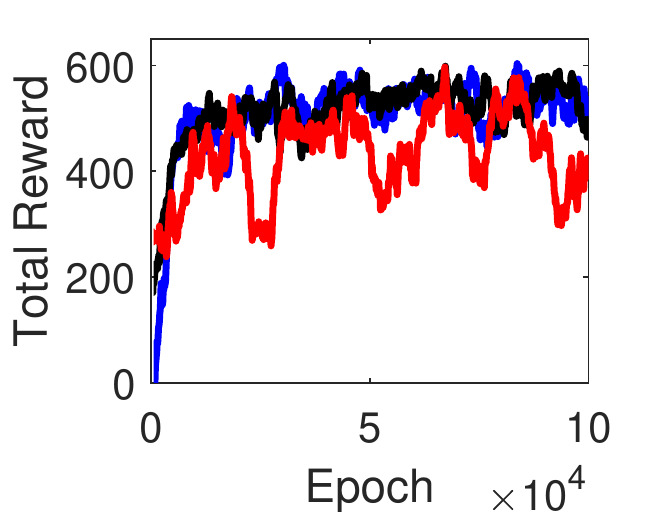} & 
\includegraphics[page=1, width=0.45\columnwidth]{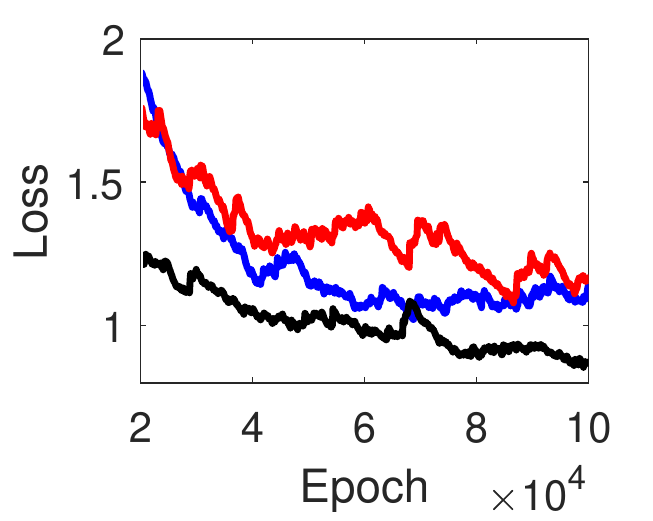} 
\tabularnewline
\tabularnewline
(a) Total Reward  & 
(b) Training Loss
\end{tabular}
\begin{tabular}{c}
\tabularnewline
\tabularnewline
\includegraphics[page=1, width=0.40\textwidth]{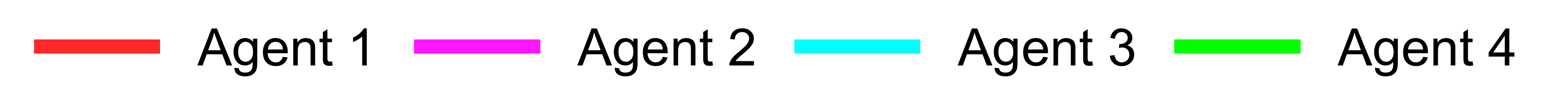}
\tabularnewline
\end{tabular}
\begin{tabular}{ccc}
\includegraphics[page=1, width=0.32\columnwidth]{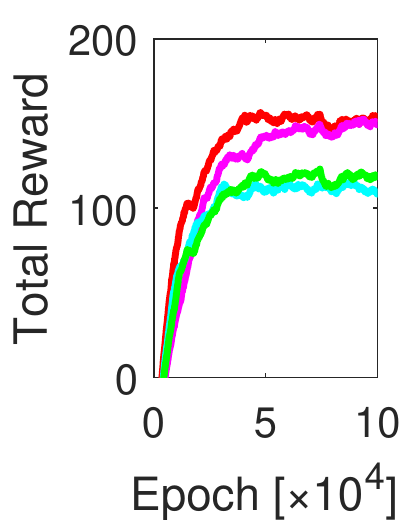} &
\includegraphics[page=1, width=0.32\columnwidth]{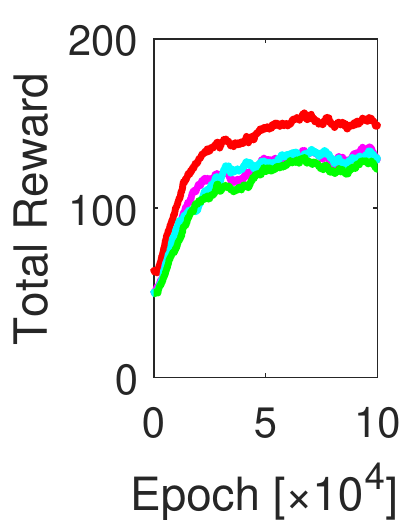} &
\includegraphics[page=1, width=0.32\columnwidth]{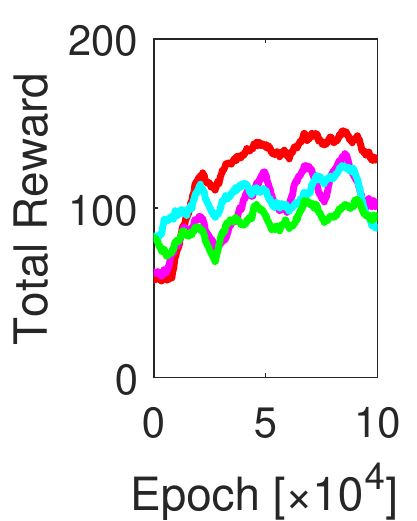} 
\tabularnewline
\tabularnewline
~~(c) Proposed & ~~~~(d) Comp1 & ~~~~(e) Comp2
\tabularnewline
\tabularnewline
\tabularnewline
\includegraphics[page=1, width=0.32\columnwidth]{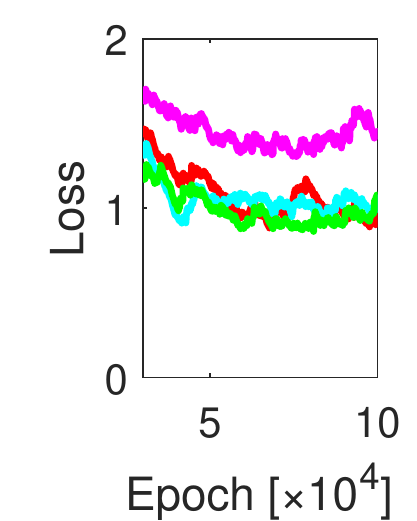} &
\includegraphics[page=1, width=0.32\columnwidth]{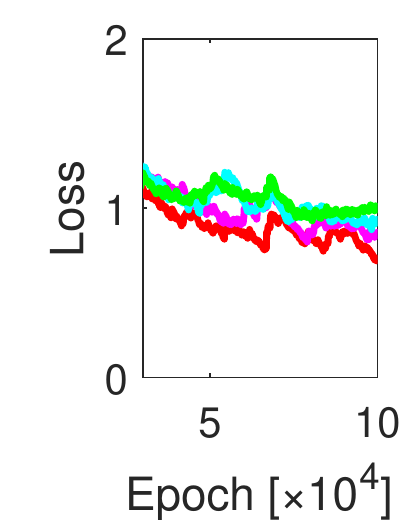} &
\includegraphics[page=1, width=0.32\columnwidth]{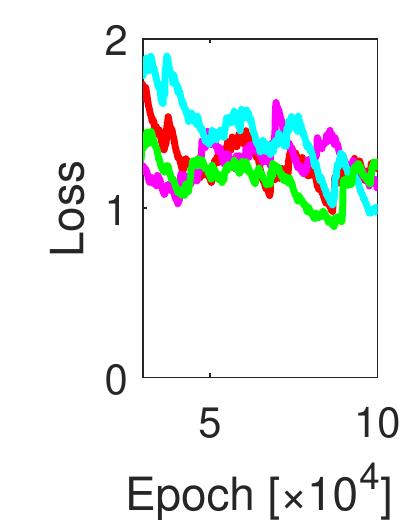} 
\tabularnewline
\tabularnewline
~~(f) Proposed & ~~~~(g) Comp1& ~~~~(h) Comp2
\end{tabular}
\caption{The learning curve of proposed scheme, Comp1, and Comp2. All agents are trained with 100,000 iterations for each scheme.}\label{fig:train}
\vspace{-10pt}
\end{figure}
\subsubsection{Reward convergence}
This work studies the tendency of training phase corresponding to total reward and training loss. Fig.~\ref{fig:train} shows the training results. As shown in Fig.~\ref{fig:train}(a)/(b), the  tendencies of two metrics in proposed scheme and Comp1 shows more stable than in Comp2. The total rewards in proposed scheme and Comp1 converge at around $[540, 580]$, whereas the total reward in Comp2 fluctuates in $[300, 580]$. The malfunction of non-agent UAVs affects the reward fluctuation due to its uncertainty. Therefore, regarding training loss, the proposed scheme starts at highest value 1.7 among the schemes and converges at 1.2 which is second rank. Comp1 and  Comp2 show the minimal and highest loss tendency, respectively. 
As shown in Fig.~\ref{fig:train}(c--e), $b_1$ commonly takes the highest reward among four agents. Fig.~\ref{fig:train}(c)/(f) show that $b_1$ and $b_2$ get high reward in proposed scheme, whereas $b_1$ and $b_2$ show the lowest and the highest loss, respectively. As shown in Fig.~\ref{fig:train}(d)/(g), agents in Comp1 show similar tendencies and smooth curves in both reward and loss. However, Comp2 shows unstable fluctuations in two metrics. To sum it up, CommNet-based decision making helps to make ACR stable with making converging the neural network to optimal in training, even though only one agent with communication operation exists as shown in Fig.~\ref{fig:train}. 
\subsubsection{Surveillance reliability} 

\begin{figure}[t!]
\centering
\setlength{\tabcolsep}{2pt}
\renewcommand{\arraystretch}{0.2}
\begin{tabular}{ccc}
\multicolumn{3}{c}{\includegraphics[page=1, width=1\columnwidth]{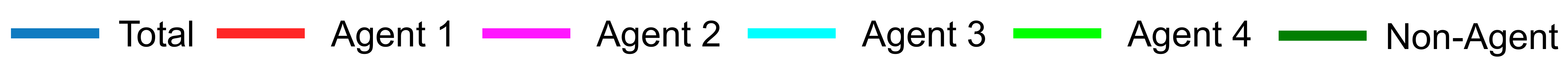}}
\tabularnewline
\includegraphics[page=1, width=0.32\columnwidth]{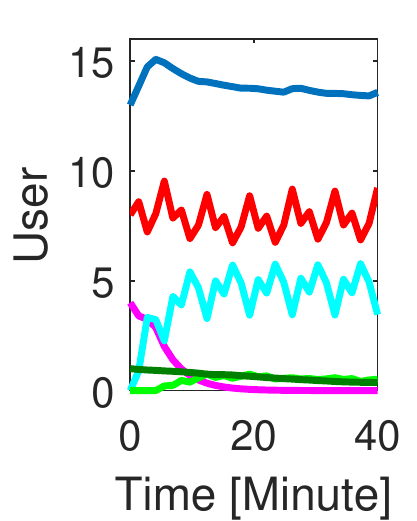} &
\includegraphics[page=1, width=0.32\columnwidth]{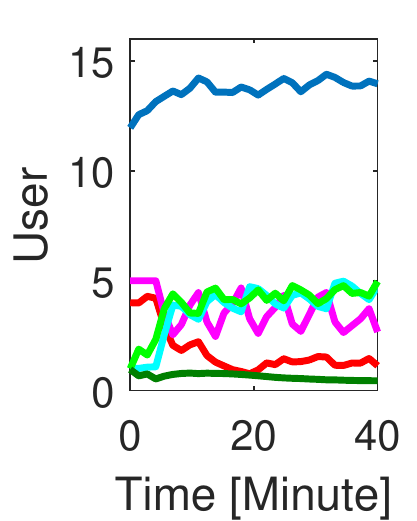} &
\includegraphics[page=1, width=0.32\columnwidth]{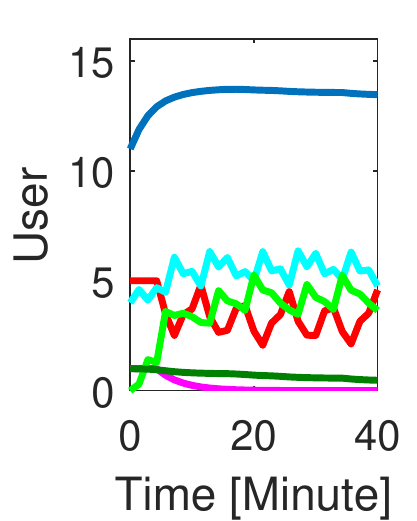}
\tabularnewline
\tabularnewline
~~(a) Proposed  & ~~~~(b) Comp1 & ~~~~(c) Comp2
\end{tabular}
\begin{tabular}{c}
\includegraphics[page=1, width=0.6\columnwidth]{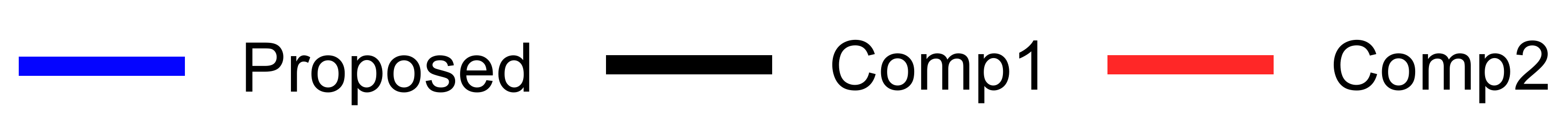}
\end{tabular}
\begin{tabular}{cc}
\includegraphics[page=1, width=0.49\columnwidth]{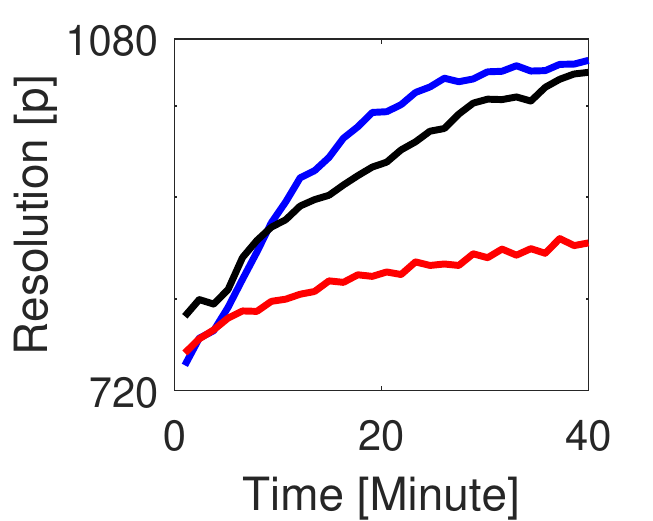} & 
\includegraphics[page=1, width=0.49\columnwidth]{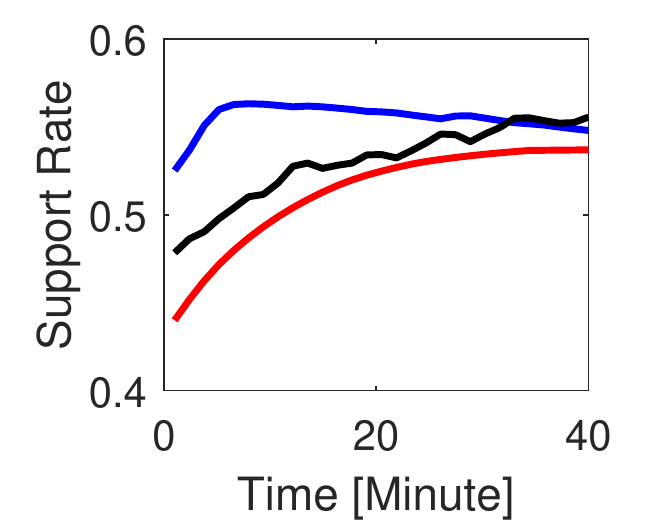} 
\tabularnewline
\tabularnewline
~~(d) Resolution  & 
~(e) Support Rate 
\end{tabular}
\caption{The performance difference of proposed scheme, Comp1, and Comp2. The graphs are plotted with trained model and taking 25 test iterations results on average.}
\label{fig:test}
\vspace{-10pt}
\end{figure}

This paper studies the impact of CommNet-based ACR on the number of users surveiled to UAV agents and surveilance resolution. The test is conducted with fixed user position and non-agent UAV position. The test results are derived with 25 test iterations. Fig.~\ref{fig:test} presents the test results with the trained model for each scheme. Fig.~\ref{fig:test}(a--c) shows the number of surveiled users. In our proposed scheme, the number of surveiled users is 12--15 on average as shown in Fig.~\ref{fig:test}(a). UAV agents  $b_1$ and $b_3$ surveil more users than other agents or 3 non-agent UAVs. In Comp1 scheme, 12--14 users are surveiled by UAVs. As shown in Fig.~\ref{fig:test}(b), all agents in Comp1 provide monitoring services to users more evenly than those of proposed or Comp2 scheme due to computing communication operations for all agents. According to 11--13 users are surveiled.
Fig.~\ref{fig:test}(d)/(e) show the resolution of monitoring service and the support rate, respectively. The surveillance resolution of all schemes starts at around $720$p.  The terminal monitoring resolution is found to be high in the order of proposed, Comp1, and Comp2. The support rate is derived from calculating the ratio of the number of surveiled user by UAV and the total number of users. The support rate is found to be high at $t=1$, in the order of the proposed scheme, Comp1, and Comp2. In addition, Comp1 is the same as the proposed scheme, and Comp2 is the lowest at $t=40$. In these results, agents in the proposed scheme take a totally different strategy. The agent that operates communication operation, surveils maximum users, whereas agent in other comparison schemes surveils users evenly. The strategy in the proposed scheme makes an outperformance in the highest resolution and maximum support rate among the benchmark scheme. 
\subsubsection{Behavior pattern analysis of UAV agents}
\begin{figure*}[t!]
\centering
\setlength{\tabcolsep}{2pt}
\renewcommand{\arraystretch}{0.2}
\begin{tabular}{c}
\includegraphics[page=1, width=0.8\textwidth]{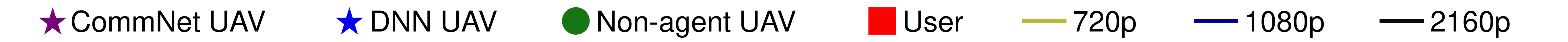}
\end{tabular}
\begin{tabular}{cccccccc}
\tabularnewline
\footnotesize ~~~~$t=5$ & 
\footnotesize ~~~~$t=10$ &
\footnotesize ~~~~$t=15$ &
\footnotesize ~~~~$t=20$ &
\footnotesize ~~~~$t=25$ &
\footnotesize ~~~~$t=30$ &
\footnotesize ~~~~$t=35$ &
\footnotesize ~~~~$t=40$  
\tabularnewline
\includegraphics[page=1, width=0.12\textwidth]{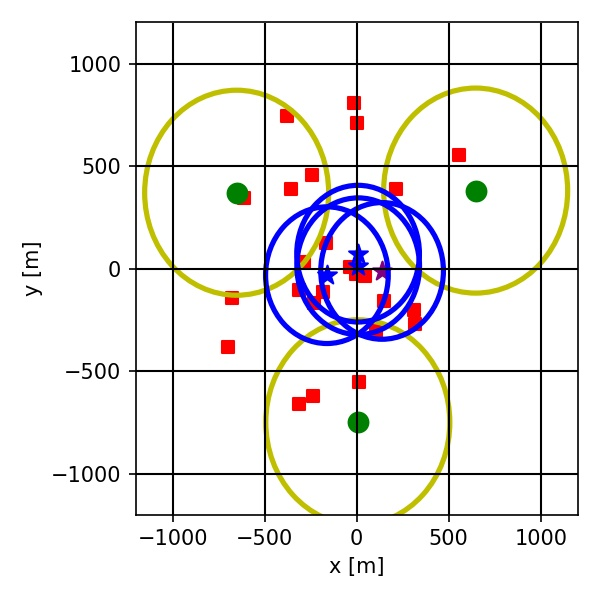} &
\includegraphics[page=1, width=0.12\textwidth]{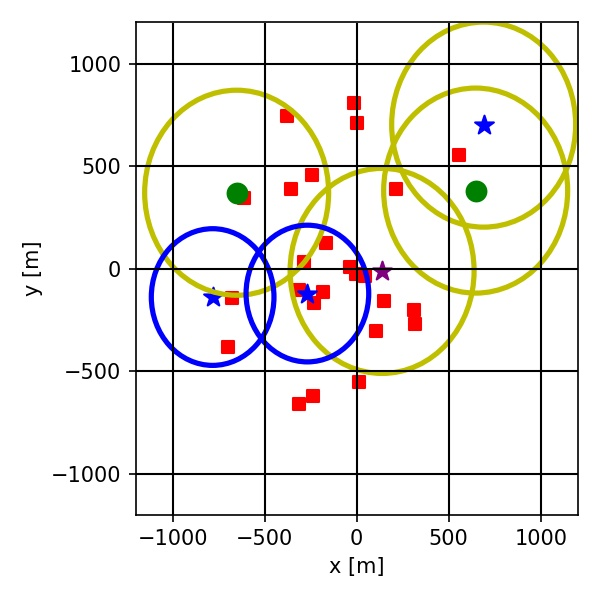} &
\includegraphics[page=1, width=0.12\textwidth]{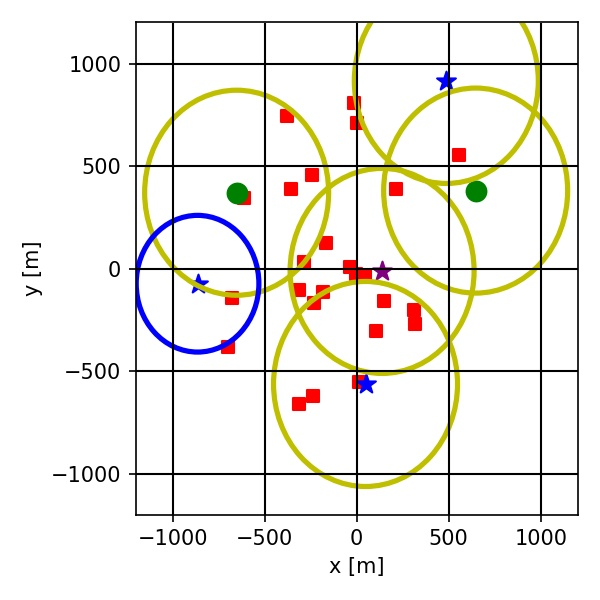} &
\includegraphics[page=1, width=0.12\textwidth]{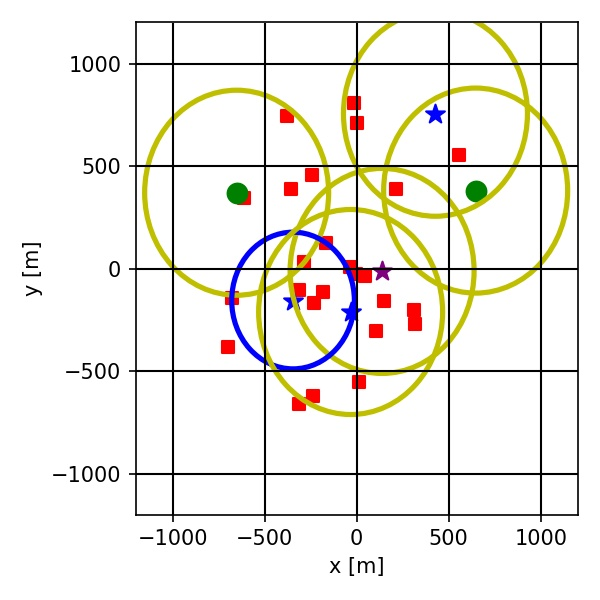} &
\includegraphics[page=1, width=0.12\textwidth]{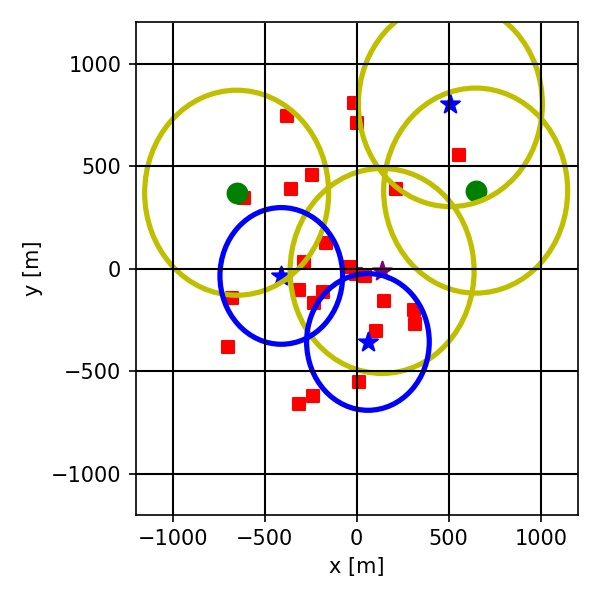} &
\includegraphics[page=1, width=0.12\textwidth]{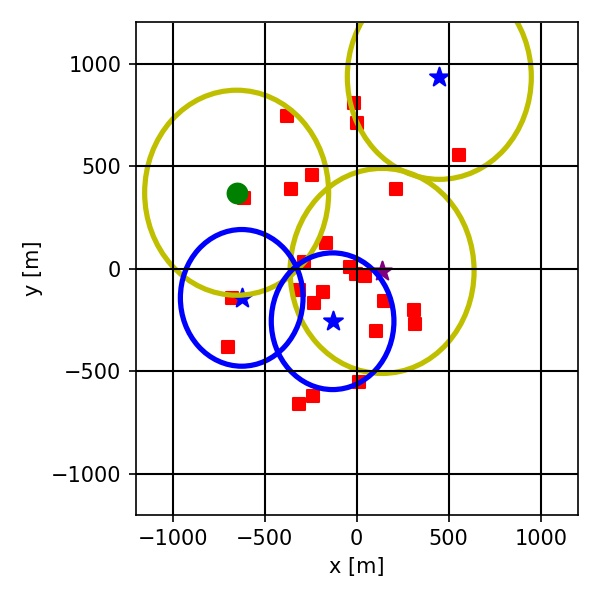} &
\includegraphics[page=1, width=0.12\textwidth]{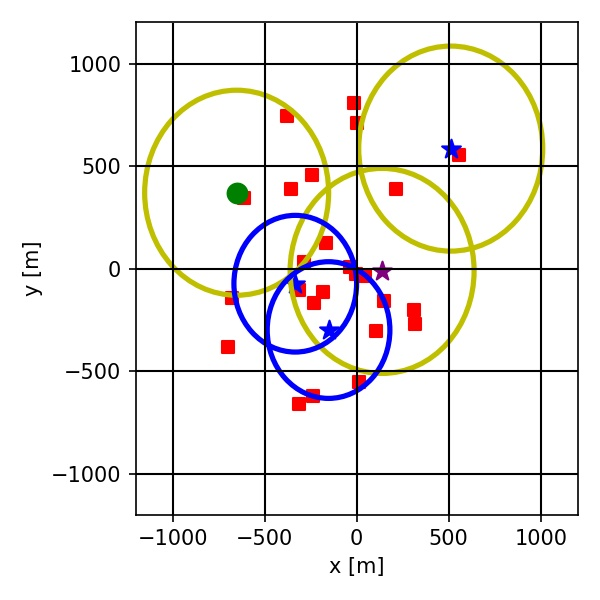} &
\includegraphics[page=1, width=0.12\textwidth]{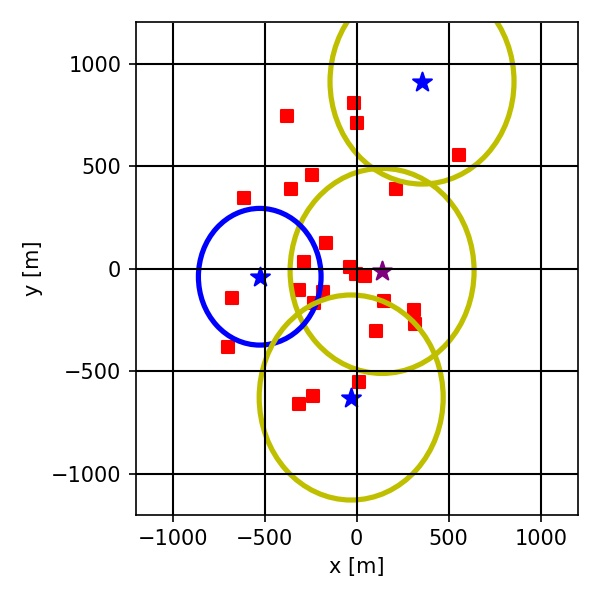} 
\tabularnewline
\tabularnewline
\multicolumn{8}{c}{(a) Behavior patterns of UAV agent in the proposed scheme}
\tabularnewline
\tabularnewline
\footnotesize ~~~~$t=5$ & 
\footnotesize ~~~~$t=10$ &
\footnotesize ~~~~$t=15$ &
\footnotesize ~~~~$t=20$ &
\footnotesize ~~~~$t=25$ &
\footnotesize ~~~~$t=30$ &
\footnotesize ~~~~$t=35$ &
\footnotesize ~~~~$t=40$ 
\tabularnewline
\includegraphics[page=1, width=0.12\textwidth]{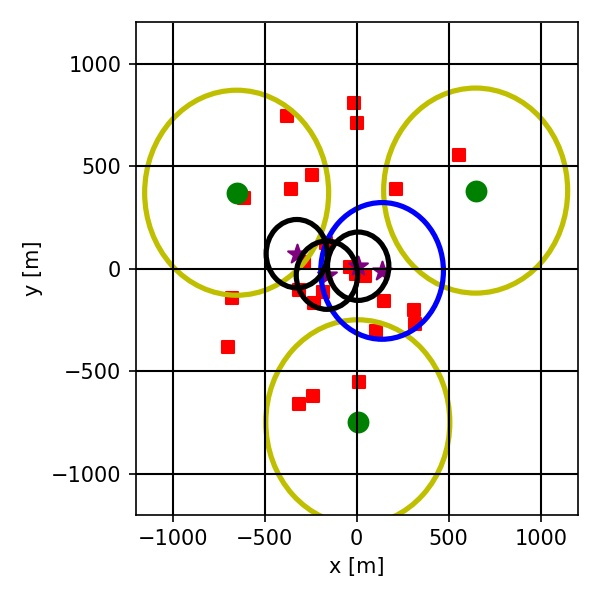} &
\includegraphics[page=1, width=0.12\textwidth]{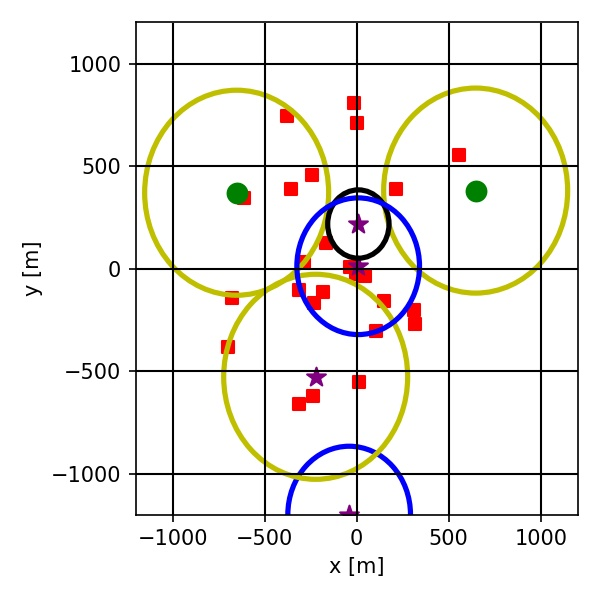} &
\includegraphics[page=1, width=0.12\textwidth]{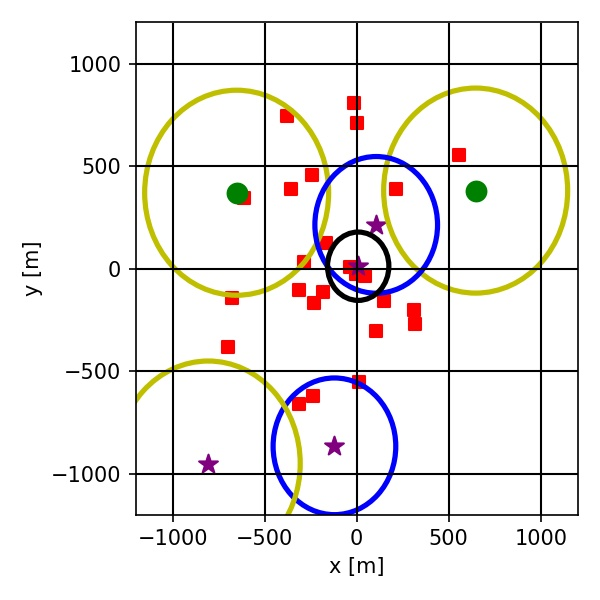} &
\includegraphics[page=1, width=0.12\textwidth]{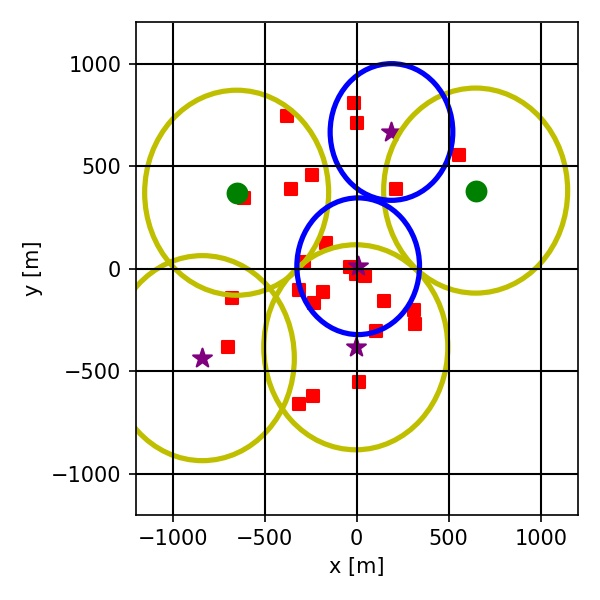} &
\includegraphics[page=1, width=0.12\textwidth]{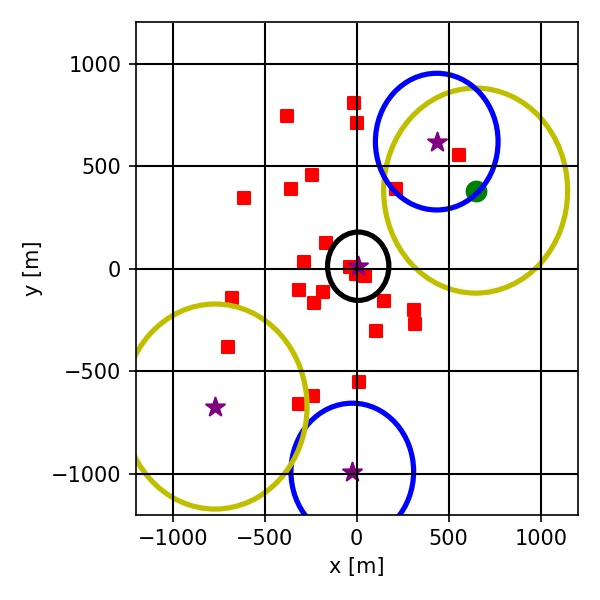} &
\includegraphics[page=1, width=0.12\textwidth]{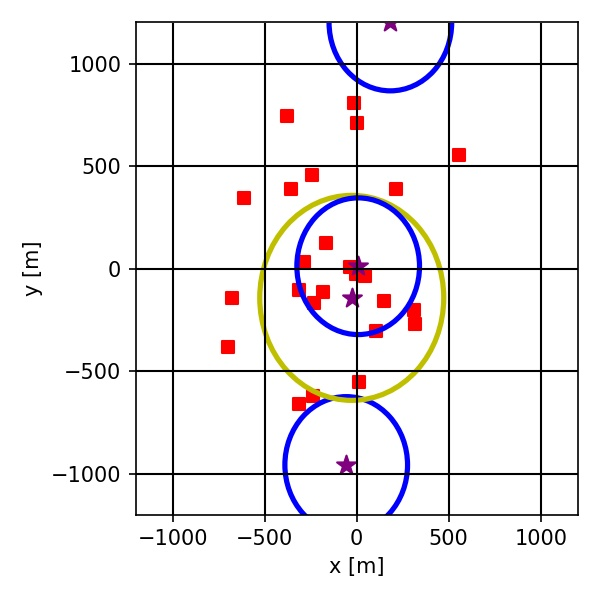} &
\includegraphics[page=1, width=0.12\textwidth]{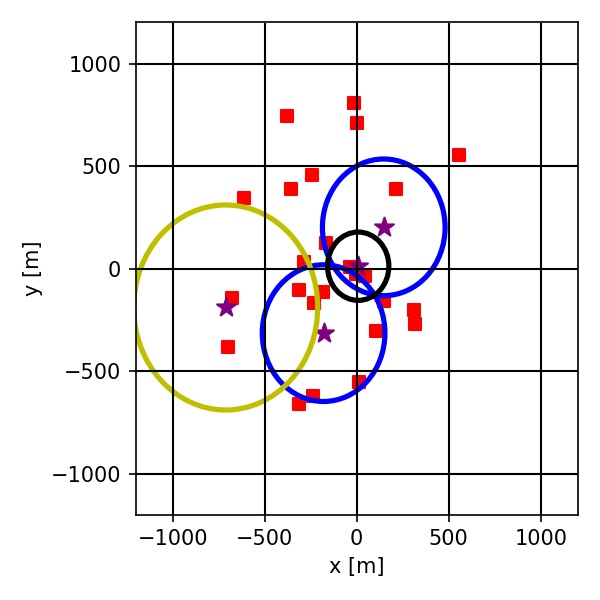} &
\includegraphics[page=1, width=0.12\textwidth]{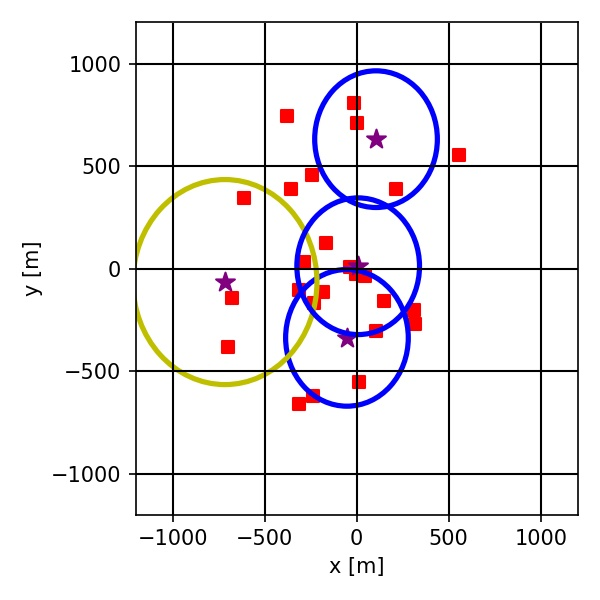} 
\tabularnewline
\tabularnewline
\multicolumn{8}{c}{(b) Behavior patterns of UAV agent in the Comp1 scheme} 
\tabularnewline
\tabularnewline
\tabularnewline
\footnotesize ~~~~$t=5$ & 
\footnotesize ~~~~$t=10$ &
\footnotesize ~~~~$t=15$ &
\footnotesize ~~~~$t=20$ &
\footnotesize ~~~~$t=25$ &
\footnotesize ~~~~$t=30$ &
\footnotesize ~~~~$t=35$ &
\footnotesize ~~~~$t=40$ 
\tabularnewline
\includegraphics[page=1, width=0.12\textwidth]{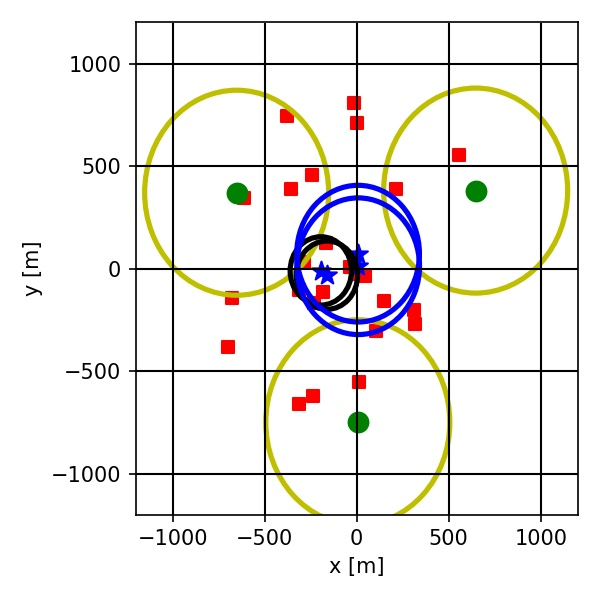} &
\includegraphics[page=1, width=0.12\textwidth]{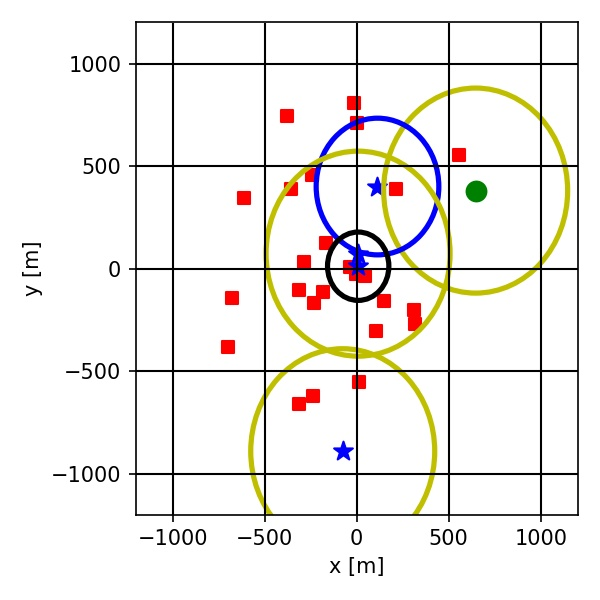} &
\includegraphics[page=1, width=0.12\textwidth]{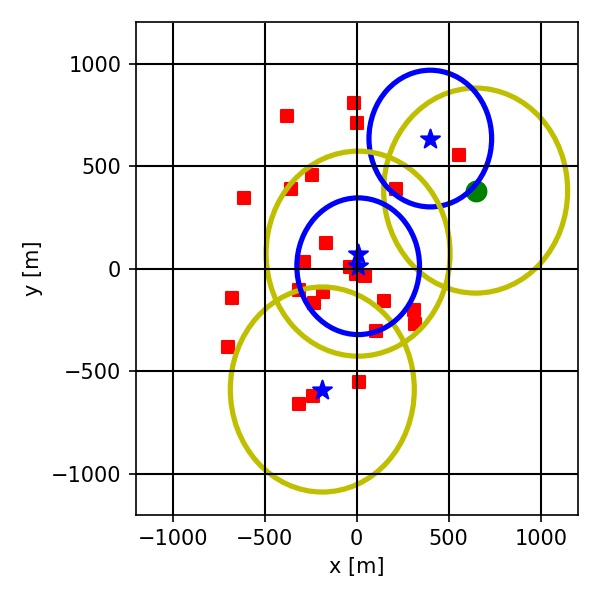} &
\includegraphics[page=1, width=0.12\textwidth]{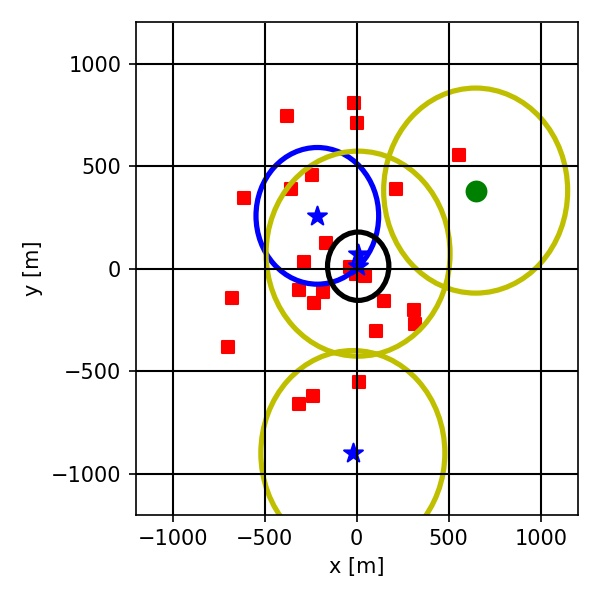} &
\includegraphics[page=1, width=0.12\textwidth]{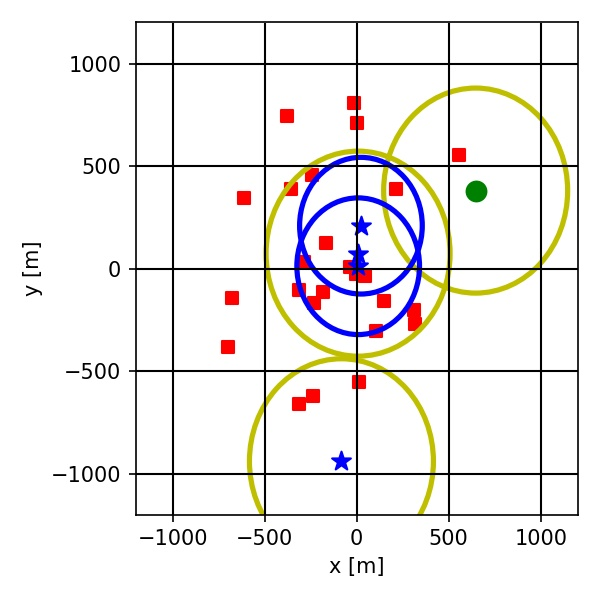} &
\includegraphics[page=1, width=0.12\textwidth]{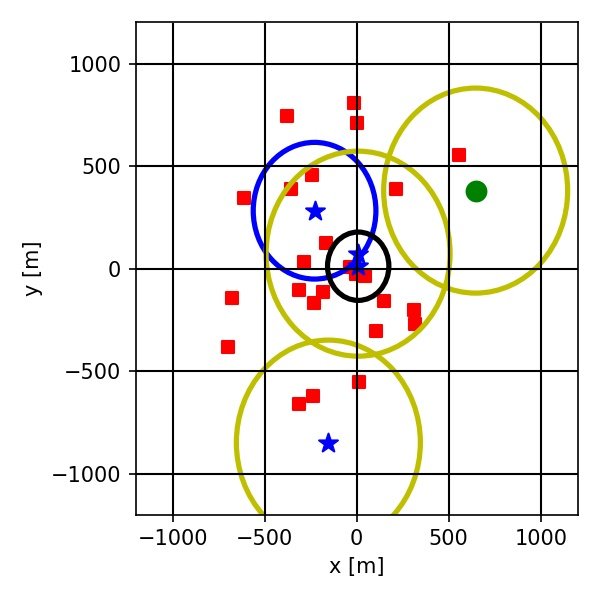} &
\includegraphics[page=1, width=0.12\textwidth]{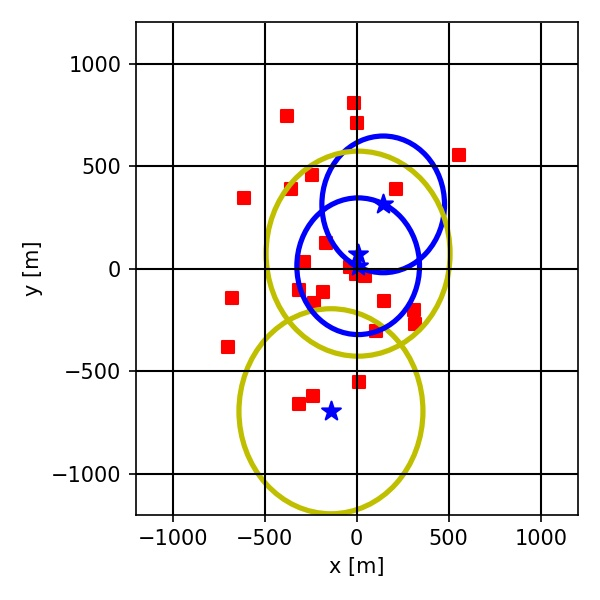} &
\includegraphics[page=1, width=0.12\textwidth]{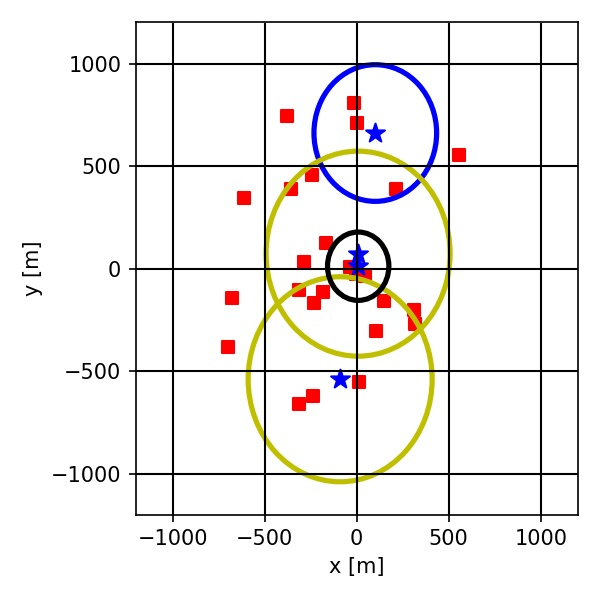}
\tabularnewline
\tabularnewline
\tabularnewline
\multicolumn{8}{c}{(c) Behavior patterns of UAV agent in the Comp2 scheme} 
\tabularnewline
\tabularnewline
\end{tabular}
 \caption{UAV agents behaviors; circle depicts the surveillance coverage range of each UAV agent, the purple/blue stars represent the position of each CommNet-based/DNN-based UAV agent, and squares stand for users, respectively. }
\label{fig:traj} 
\end{figure*}
Fig.~\ref{fig:traj} shows how the trained UAV agents adjust the optimal location and surveillance coverage over time. Fig.~\ref{fig:traj}(a) shows the trajectories and coverage of UAVs in the proposed scheme, Fig.~\ref{fig:traj}(b)/(c) present the trajectories and coverage of UAVs in Comp1 and Comp2, respectively. Four agents consist of one CommNet UAV and three DNN UAVs for the proposed scheme, four CommNet UAVs for the Comp1 scheme, and four DNN UAVs for the Comp2 scheme.
As shown in Fig.~\ref{fig:traj}(a), the non-agent UAVs are malfunctioned at $t\in(5,10]$, $t\in(25,30]$, and $t\in(35,40]$, respectively. The UAV agent with CommNet-based policy, occupies the optimal position where the most users located at $t\in(5,40]$. The UAV agents with DNN-based policy, prowl to surveil the remaining users while enhancing the resolution. The strategy in the proposed scheme makes the fastest increasing of resolution as shown Fig.~\ref{fig:test}(d). Fig.~\ref{fig:traj}(b) presents Comp1 scheme that consists of all agents use CommNet-based policies. The malfunctioned events occur at $t\in(5,10]$, $t\in(20,25]$, and $t\in(25,30]$. There is static CommNet agent that controls the resolution between 1080p and 2160p. Two agents moves to where the malfunctioned event occurs at $t \in (5, 10]$. All users get monitoring service at $t = 20$. When $t \in (20,40]$, all UAV agents mosey to guarantee surveillance resolution and the number of users. As shown in Fig.~\ref{fig:test}(d)/(e), The strategy of Comp1 increases the resolution and support rate even though all non-agent UAVs are malfunctioned. Fig.~\ref{fig:traj}(c) shows the behavior pattern of DNN-based UAV agents in Comp2 scheme. The non-agent UAVs are malfunctioned at $t\in(5,10]$, $t\in(30,35]$, and $t\in(35,40]$. Comp2 shows the highest overlapped area. Agents in Comp2 try to occupy preempt an area from the other agents at $t\in[10,40)$. This behavior pattern causes the lowest performance on every performance metrics, including not only benchmark of training and test as shown in Fig.~\ref{fig:train} and Fig.~\ref{fig:test}.
\subsubsection{Computational cost and efficiency} 
\begin{table}[t!] 
\scriptsize
\small
\caption{Comparison of computational cost between \textit{CommNet-based} and \textit{DNN-based} policy and their application scheme (Proposed, Comp1, and Comp2).}
\centering
\begin{tabular}{c|l||c}\toprule
\multicolumn{2}{c||}{\textbf{Metric}} & \textbf{Computational Cost [FLOPS]}
\tabularnewline\midrule 
\multirow{2}{*}{\textbf{Policy}}&CommNet&  $222,521$
\tabularnewline
& DNN &  $181,369$
\tabularnewline\midrule
\multirow{3}{*}{\textbf{Scheme}}& Proposed & $766,628$
\tabularnewline
& Comp1 & $890,084$
\tabularnewline
& Comp2 & $725,476$
\tabularnewline\toprule
\end{tabular}
\label{tab:cost}
\vspace{-10pt}
\end{table}

This study looks at the computational cost of CommNet-based policy and DNN-based policy. Furthermore, this paper figures out the total computational cost of our proposed scheme, Comp1, and Comp2. 
According to \cite{justus2018predicting}, the computational cost of the neural network can be calculated. As shown in Table~\ref{tab:cost}, the computational cost of {CommNet}-based policy is 22.7\% larger than {DNN}-based policy. In the proposed scheme, there is a leader UAV agent. In other words, a leader UAV requires CommNet-based policy to collects the observations of other agents. All agents have CommNet-based policies for Comp1 and DNN-based policies for Comp2. Regarding computational cost for each time step, the proposed scheme requires 16.1\% fewer FLOPS than Comp1, and 5.3\% more FLOPS than Comp2. 
Even Comp1 requires higher computational cost than other schemes, and our proposed scheme shows stable convergence and high performance as much as Comp1. The proposed scheme requires a higher computational cost than Comp2 and shows performance superiority to Comp2. Therefore, our proposed scheme outperforms other comparison schemes.
\subsection{Discussion} This subsection discusses the reason why these experimental results are derived. According to \cite{CommNet,foerster2016learning}, no communication between an inter-agent does not guarantee collaboration for surveillance reliability. In the proposed scheme and Comp1, the cooperative reward enables the cooperation between an inter-UAV. In Comp2, no communication between UAVs leads to a lack of cooperation due to the non-communicative neural network structure.
In addition, it is remarkable that the performance of ACR increases even though only one CommNet UAV agent exists. Due to this aspect, our proposed scheme shows the equal or more extraordinary performance of Comp2 with less computational cost.

\section{Conclusions and Future Work}\label{sec:con}
{This paper considers} the deployment of surveillance UAVs, which aims at energy-efficient and reliable surveillance. Using UAVs to improve monitoring services and handle various uncertainty problems, this paper proposes an autonomous surveillance UAVs cooperation scheme based on cooperative model-free MADRL, called CommNet. 
The proposed scheme offers a promising and reliable solution to find optimal trajectories in the operating area and surveillance coverage control of UAVs that can cover as many users as possible. In addition, the proposed scheme outperforms comparison techniques in the computational efficiency as well as the benchmarks. The future work of this study is to expand ACR under various conditions (i.e., ACR with geographic constraints~\cite{future}) and diverse applications (i.e., federated learning applications to ACR in order to train data those are collected from different agents).
\bibliographystyle{IEEEtran}

\begin{IEEEbiography}[{\includegraphics[width=1in,height=1.25in,clip]{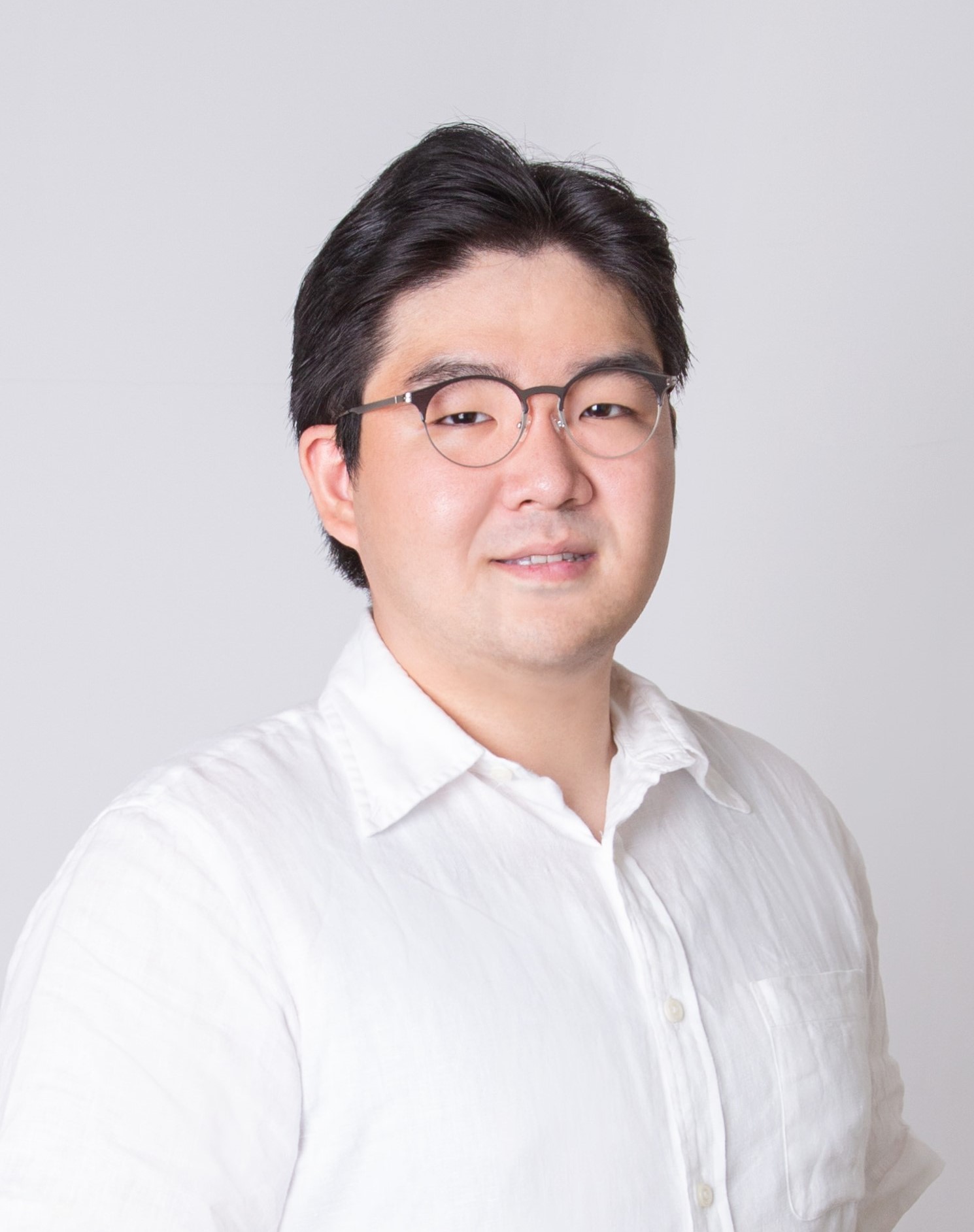}}]{Won Joon Yun} is currently a Ph.D. student in electrical and computer engineering at Korea University, Seoul, Republic of Korea, since March 2021, where he received his B.S. in electrical engineering. His current research interests include multi-agent deep reinforcement learning for various mobile and network systems. He was a recipient of the Best Paper Awards by KICS (2020--2021) and IEEE ICOIN Best Paper Award (2021).
\vspace{-10pt}
\end{IEEEbiography}

\begin{IEEEbiography}[{\includegraphics[width=1in,height=1.25in,clip]{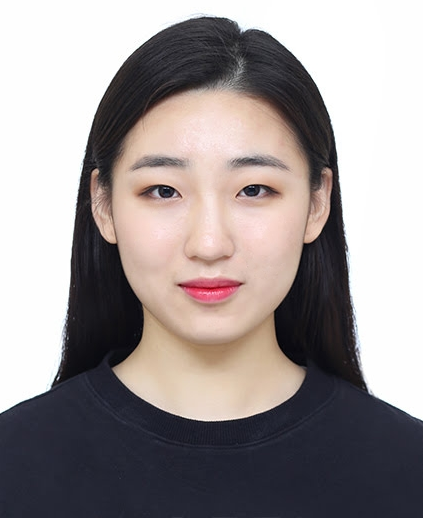}}]{Soohyun Park} is currently pursuing the Ph.D. degree in electrical engineering with Korea University, Seoul, Republic of Korea. She received her B.S. in computer science and engineering from Chung-Ang University, Seoul, Republic of Korea, in 2019. Her research interest includes deep learning algorithms and their applications. She was a recipient of the IEEE Vehicular Technology Society (VTS) Seoul Chapter Award, in 2019.
\vspace{-10pt}
\end{IEEEbiography}

\begin{IEEEbiography}[{\includegraphics[width=1in,height=1.25in,clip]{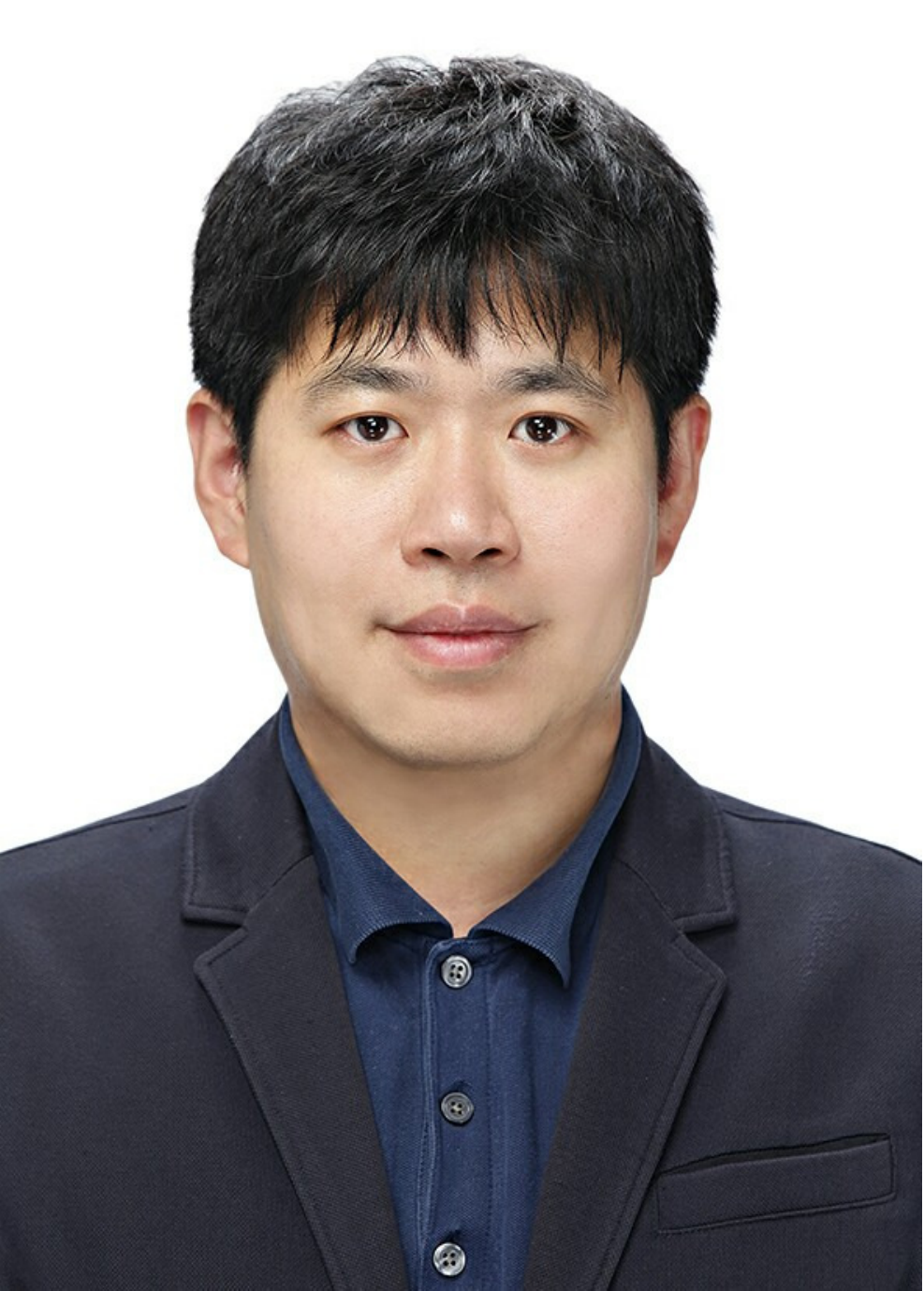}}]{Joongheon Kim}
(Senior Member, IEEE) has been with Korea University, Seoul, Korea, since September 2019, and he is currently an associate professor at the School of Electrical Engineering. He is also a vice director of the Artificial Intelligence Engineering Research Center at Korea University, Seoul, Korea. 
He received the B.S. and M.S. degrees in computer science and engineering from Korea University, Seoul, Korea, in 2004 and 2006, respectively; and the Ph.D. degree in computer science from the University of Southern California (USC), Los Angeles, California, USA, in 2014. 
Before joining Korea University, he was with LG Electronics CTO Office, Seoul, Korea, from 2006 to 2009; InterDigital, San Diego, California, USA, in 2012; Intel Corporation, Santa Clara in Silicon Valley, California, USA, from 2013 to 2016; and Chung-Ang University, Seoul, Korea, from 2016 to 2019. 

He is a senior member of the IEEE and serves as an associate/guest editor for \textit{IEEE Transactions on Vehicular Technology}, \textit{IEEE Communications Standards Magazine}, and \textit{Computer Networks (Elsevier)}. He is also a distinguished lecturer for \textit{IEEE Communications Society} (2022--2023).
He was a recipient of the Annenberg Graduate Fellowship with his Ph.D. admission from USC (2009), 
Intel Corporation Next Generation and Standards (NGS) Division Recognition Award (2015), Haedong Young Scholar Award by KICS (2018), Paper Awards from IEEE Seoul Section Student Paper Contests (2019 and 2020), \textit{IEEE Systems Journal} Best Paper Award (2020), IEEE ICOIN Best Paper Award (2021), Haedong Paper Award by KICS (2021), and IEEE Vehicular Technology Society (VTS) Seoul Chapter Awards (2019 and 2021).
\vspace{-10pt}
\end{IEEEbiography}

\begin{IEEEbiography}[{\includegraphics[width=1in,height=1.25in,clip]{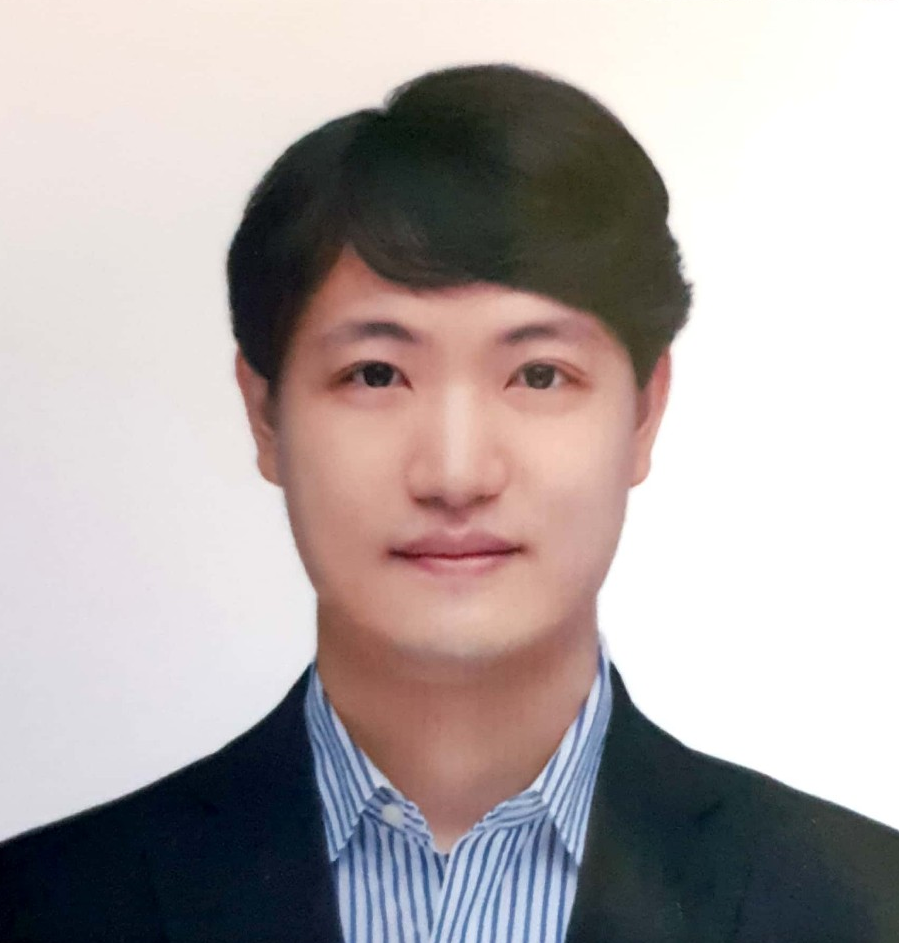}}]{MyungJae Shin} is currently an AI researcher at Mofl Inc., Daejeon, Republic of Korea.
He received his B.S. (\textit{second highest honor} from the College of Engineering) and M.S. degrees in computer science and engineering from Chung-Ang University, Seoul, Korea, in 2018 and 2020. 

His research interests are in various econometric theories and their deep-learning based computational solutions. He was a recipient of National Science \& Technology Scholarship (2016--2017) and IEEE Vehicular Technology Society (VTS) Seoul Chapter Award (2019).
\vspace{-10pt}
\end{IEEEbiography}
	
\begin{IEEEbiography}[{\includegraphics[width=1in,height=1.25in,clip]{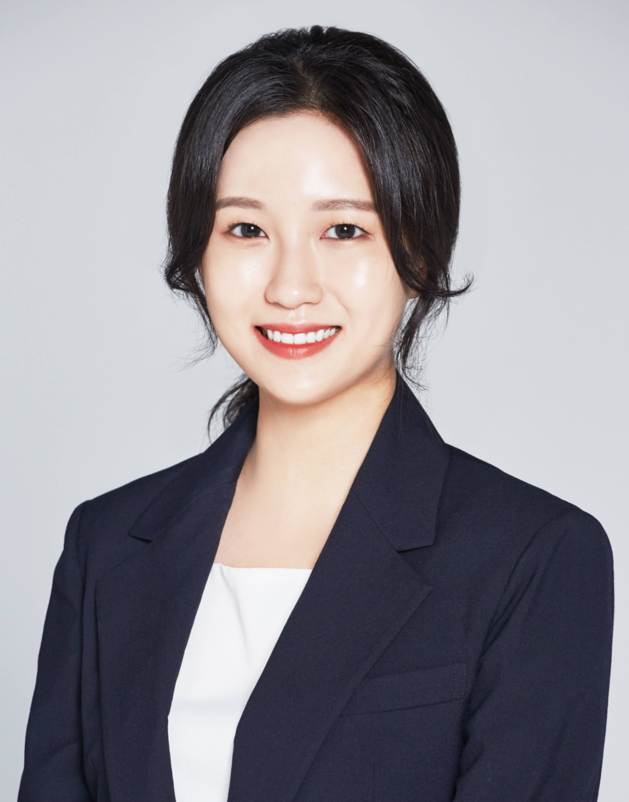}}]{Soyi Jung} (Member, IEEE) has been an assistant professor at the School of Software, Hallym University, Chuncheon, Republic of Korea, since September 2021. She also holds a visiting scholar position at Donald Bren School of Information and Computer Sciences, University of California, Irvine, CA, USA, from 2021 to 2022. She was a research professor at Korea University, Seoul, Republic of Korea, during 2021. She was also a researcher at Korea Testing and Research (KTR) Institute, Gwacheon, Republic of Korea, from 2015 to 2016. 
She received her B.S., M.S., and Ph.D. degrees in electrical and computer engineering from Ajou University, Suwon, Republic of Korea, in 2013, 2015, and 2021, respectively. 

Her current research interests include network optimization for autonomous vehicles communications, distributed system analysis, big-data processing platforms, and probabilistic access analysis. She was a recipient of Best Paper Award by KICS (2015), Young Women Researcher Award by WISET and KICS (2015), Bronze Paper Award from IEEE Seoul Section Student Paper Contest (2018), ICT Paper Contest Award by Electronic Times (2019), and IEEE ICOIN Best Paper Award (2021).
\vspace{-10pt}
\end{IEEEbiography}

\begin{IEEEbiography}[{\includegraphics[width=1in,height=1.25in,clip]{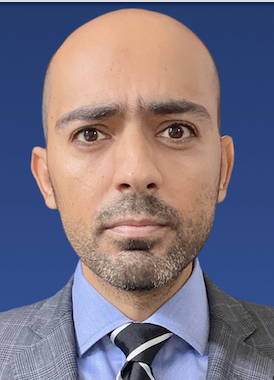}}]{David Mohaisen}
received the M.Sc. and Ph.D. degrees from the University of Minnesota, Minneapolis, MN, USA, in 2012. He was an Assistant Professor with SUNY Buffalo, Buffalo, NY, USA, from 2015 to 2017 and a Senior Research Scientist with Verisign Labs, Reston,
VA, USA, from 2012 to 2015. In 2017, he joined the University of Central Florida, Orlando, FL, USA, as an Associate Professor, where he only directs the Security and Analytics Laboratory. His research interests are in the areas of networked systems and their security, online privacy, and measurements. Dr. Mohaisen is an Associate Editor of \textit{ACM Transactions on Mobile Computing} and a Senior Member of ACM in 2018.
\vspace{-10pt}
\end{IEEEbiography}

\begin{IEEEbiography}[{\includegraphics[width=1in,height=1.25in,clip]{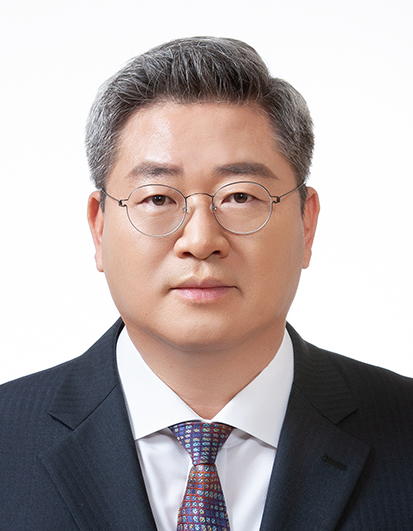}}]{Jae-Hyun Kim} received the B.S., M.S., and Ph.D. degrees, all in computer science and engineering, from Hanyang University, Ansan, Korea, in 1991, 1993, and 1996 respectively. In 1996, he was with the Communication Research Laboratory, Tokyo, Japan, as a Visiting Scholar. From April 1997 to October 1998, he was a postdoctoral fellow at the department of electrical engineering, University of California, Los Angeles. From November 1998 to February 2003, he worked as a member of technical staff in Performance Modeling and QoS management department, Bell laboratories, Lucent Technologies, Holmdel, NJ. He has been with the department of electrical and computer engineering, Ajou University, Suwon, Korea, as a professor since 2003. 

He is the Center Chief of Satellite Information Convergence Application Services Research Center (SICAS) sponsored by Institute for Information $\&$ Communications Technology Promotion in Korea. He is Chairman of the Smart City Committee of 5G Forum in Korea since 2018. He is Executive Director of the Korea Institute of Communication and Information Sciences (KICS). He is a member of the IEEE, KICS, the Institute of Electronics and Information Engineers (IEIE), and the Korean Institute of Information Scientists and Engineers (KIISE). He was a recipient of IEEE ICOIN Best Paper Award (2021).
\vspace{-10pt}
\end{IEEEbiography}
\end{document}